\documentclass[preprint]{aastex}
\usepackage{amsbsy} 

\newcommand{\Msun}{M_\odot}
 
\newcommand{\bh}{\boldsymbol{h}}
 
\newcommand{\bx}{\boldsymbol{x}}
\newcommand{\by}{\boldsymbol{y}}

\slugcomment{submitted to MNRAS}

\shorttitle{The Dynamical Evolution of the Pleiades}
\shortauthors{Converse \& Stahler}

\begin{document}

\title{The Dynamical Evolution of the Pleiades}

\author{Joseph M. Converse and Steven W. Stahler}

\affil{Astronomy Department. University of California,
Berkeley, CA 94720}

\email{jconverse@astro.berkeley.edu}

\begin{abstract}
We present the results of a numerical simulation of the history and future
development of the Pleiades. This study builds on our previous one that 
established statistically the present-day structure of this system. Our 
simulation begins just after molecular cloud gas has been expelled by the 
embedded stars. We then follow, using an N-body code, the stellar dynamical 
evolution of the cluster to the present and beyond. Our initial state is that 
which evolves, over the 125 Myr age of the cluster, to a configuration most 
closely matching the current one.

We find that the original cluster, newly stripped of gas, already had a virial
radius of 4~pc. This configuration was larger than most observed, embedded 
clusters. Over time, the cluster expanded further and the central surface 
density fell by about a factor of two. We attribute both effects to the 
liberation of energy from tightening binaries of short period. Indeed, the 
original binary fraction was close to unity. The ancient Pleiades also had 
significant mass segregation, which persists in the cluster today. 

In the future, the central density of the Pleiades will continue to fall. For 
the first few hundred Myr, the cluster as a whole will expand because of
dynamical heating by binaries. The expansion process is aided by mass loss
through stellar evolution, which weakens the system's gravitational binding.
At later times, the Galactic tidal field begins to heavily deplete the cluster
mass. It is believed that most open clusters are eventually destroyed by
close passage of a giant molecular cloud. Barring that eventuality, the
density falloff will continue for as long as 1~Gyr, by which time most of the
cluster mass will have been tidally stripped away by the Galactic field. 
\end{abstract}


\section{Introduction}

Despite recent advances in the field of star formation, the origin of open
clusters remains a mystery. It is now generally accepted that all stars are 
born within groups. These groups are at first heavily embedded within molecular
clouds, their members obscured optically by copious interstellar dust. By the
time the stars are revealed, only about 10~percent are in open clusters
\citep{ms78,am01}. The remainder are in either T- or OB associations, both
destined to disperse within a few Myr. In contrast, the stars within open
clusters are gravitationally bound to each other, and the group can survive 
intact for several Gyr \citep{f95}. How do molecular clouds spawn these 
relatively rare but stable configurations?

One intriguing aspect of the mystery is that open clusters are intermediate in
their properties between T- and OB associations. The former are relatively 
sparse in projected stellar density, and contain up to about 100 members
\citep[e.g.,][]{kh95,l07}. The latter, as exemplified by the nearby Orion 
Nebula Cluster, begin with extraordinarily high density \citep{ms94} and 
contain well over a thousand members \citep{h97}, far more than the eponymous 
O and B stars. A published compilation of Galactic open clusters \citep{m95} 
shows them to have from a few hundred to roughly a thousand stars, i.e., just
in the middle range. Apparently, systems born with either too low or too high 
a population and density are fragile, while the relative minority falling in 
between can survive.

There is already an extensive literature on young, bound clusters, both 
observational and theoretical \citep[for a review, see][]{e00}. Models for 
their origin, dating back at least to \citet{l84}, have focused on the need for
a high star formation efficiency in the parent cloud. A standard computational 
technique, using N-body simulations, is to create stars in a background 
potential well, remove that potential through various prescriptions, and then 
assess the result \citep[e.g.,][]{gb06,bk07}. Some researchers using this  
approach, implemented either analytically or numerically,  have hypothesized 
that open clusters are the bound remnants of expanding OB~associations 
\citep{a00,k01}. In recent years, most theoretical ideas have been motivated 
by fluid dynamical simulations of turbulent, collapsing clouds 
\citep{k00,v03,l03}. While much insight has been gained from these collective 
investigations, there has generally been too little contact of the theory with 
actual groups. A clear advance would be made if we could establish empirically 
the original state of one or more observed clusters. We would then be in a 
position to gauge how these particular systems were produced by star formation 
activity in their parent clouds. 

As a first step in this direction, \citet[][hereafter Paper I]{cs08} undertook
a quantitative study of the {\it present-day} structure of a well-studied, 
relatively nearby open cluster, the Pleiades. We derived statistically, using a
maximum likelihood analysis, such key properties as the stellar density 
distribution, mass function, overall binary fraction, and correlation between 
the component masses of these binaries. The point of that study was to provide 
the endpoint for any calculation of the system's previous evolution.

We now take the second step. The age of the Pleiades has been determined, from
observations of lithium depletion, to be 125~Myr \citep{s98}. Using the
publicly available code Starlab \citep[Appendix B]{pz01}, we 
have run a suite of N-body calculations over just this time period, to find 
that initial state which evolved to the current cluster, as gauged by our 
previous investigation. In doing so, we also establish the detailed history of 
the group over that epoch, and even into the future.

A key assumption here is that the Pleiades divested itself of cloud gas 
relatively soon after its birth. There is currently no direct means to assess 
the duration of the initial, embedded phase, either in the Pleiades or any 
other open cluster We may take a clue from T~associations, which are still 
surrounded (but not completely obscured) by molecular gas. No systems are 
observed with ages exceeding about 5~Myr, a striking fact first noted by 
\citet{h78}. Presumably, older groups consisting of post-T~Tauri stars have 
already driven away their clouds and are merged observationally into the field 
population. If a similar embedded period held for the Pleiades, it indeed 
represents a small fraction of the total age. Hence, we can establish, with 
some confidence, the cluster's structure just after cloud dispersal. A future 
study will investigate, using a combination of gaseous and stellar dynamics, 
how this early configuration itself arose. 

In Section~2 below, we describe in more detail our approach to the problem. We
define the parameters characterizing both the initial configuration of 
the cluster and the evolved system. We then outline our strategy for finding 
the optimal initial state, i.e., the one whose descendent matches most closely 
the current Pleiades. Our actual numerical results are presented in Section~3. 
Here, we give the detailed properties of the inferred initial state. We also 
describe how the cluster changed up to the present, and how it will develop in 
the future. One of our key findings is that the cluster's evolution did 
{\it not} proceed in the classic manner associated with dynamical relaxation 
\citep[][Chapter~8]{bt87}; we explore the origin of this discrepancy. Finally, 
Section~4 discusses the implications of our findings on the earlier, embedded 
evolution of this, and other, open clusters.  

\section{Method of Solution}
\subsection{Initial Cluster Parameters}
\subsubsection{Density and Velocity Distribution}

As in Paper I, we model the Pleiades as a perfectly spherical system, although
the cluster is observed to be slightly elongated \citep{rm98}. This elongation
seems to have been created by the tidal gravity of the Galaxy \citep{w74},
which would have exerted influence throughout the cluster's dynamical history.
We assume that this modest tidal stretching had negligible effect on the
internal evolution, and that relatively few stars were lost by tidal stripping
over the Pleiades age. Thus we can safely ignore the associated Galactic 
potential. Similarly, we ignore mass loss through stellar evolution, which is 
negligible for our adopted mass function, over the 125 Myr age of the 
Pleiades. In Sections 3.2 and 3.3, we present simulations that include both 
effects

Returning to our standard runs, we further assume that the cluster was in 
virial equilibrium following expulsion of the gas. Any significant departure 
from equilibrium would be erased on a dynamical time scale, about 10~Myr for 
our input parameters and therefore much shorter than the evolutionary span of 
interest. Two popular choices for spherical equilibria are \citet{k66} models 
and polytropes. As will be explained in Section 2.3, King models do not include
low enough density contrasts for a full exploration of initial states. We 
therefore used polytropes, which are more versatile in this regard. In 
polytropes, the stellar distribution function $f$, i.e., the number of stars 
per volume in configuration and velocity space, is given by
\begin{equation}
f ({\cal E}) = \left\{
\begin{array}{rl}
  A\,{\cal E}^{n-3/2} & \,\,{\cal E}\,\ge\,0\\
  0                   & \,\,{\cal E}\,<\,0\,\,.
\end{array} \right.
\end{equation}
Here, $A$ is a normalization constant, while $\cal E$, the relative energy per 
unit mass, is 
\begin{equation}
{\cal E} \,\equiv\, \Psi (r) \,-\, v^2/2 \,\,. 
\end{equation}
In this last equation, $v$ denotes the stellar speed. Thus, $\cal E$ is the
negative of the physical energy, and also has an offset in its zero point, as
conventionally defined. This offset is embedded in the relative potential 
$\Psi (r)$, which is related to the usual gravitational potential $\Phi (r)$ 
by
\begin{equation}
\Psi (r) \,\equiv\, \Phi (r_t) \,- \, \Phi (r) \,\,.
\end{equation}
The tidal radius $r_t$ marks the outer boundary reached by cluster stars. By
construction, the relative potential $\Psi$ is positive inside the cluster and 
falls to zero at \hbox{$r\,=\,r_t$}.  

The number of stars per unit volume is found by integrating the distribution 
function over velocity space. The manipulations here are standard
\citep[][Chapter 4]{bt87}, so we give only the essential results. We let $m$
denote the stellar mass, assumed provisionally to be identical for all cluster
members. Then $\rho$, the mass density of stars, is
\begin{equation}
\rho (r) \,=\, 4\,\pi\,m\,A \,\int_0^{v_{\rm max}}\! {\cal E}^{n-3/2}\,v^2\,dv
\,\,. 
\end{equation} 
Here, \hbox{$v_{\rm max}(r)\,\equiv\, \sqrt{2\,\Psi (r)}$} is the maximum speed
for a star at radius $r$. For such a star, \hbox{${\cal E}\,=\,0$}. The total
physical energy per unit mass, \hbox{$\Phi (r)\,+\, v^2/2$}, is $\Phi (r_t)$,
so the star can just reach $r_t$. Using equation~(2), equation~(4) becomes 
\begin{equation}
\rho (r) \,=\, 4\,\pi\,m\,A\,\Psi^{n-3/2}\,\int_0^{\sqrt{2\Psi}}\!
\left(1 \,-\,{v^2\over{2\,\Psi}}\right)^{n-3/2}
\,v^2\,dv \,\,.
\end{equation}
We define a new variable
\hbox{$\theta\,\equiv\,{\rm arcsin}\,(v/\sqrt{2\Psi})$}, so that
\begin{eqnarray}
\rho (r) &=& 2^{5/2}\,\pi\,m\,A\, {\Psi^n\over n} \,
\int_0^{\pi/2}\!{\rm cos}^{2n-2}\,\theta\,d\theta  \\
     &=& (2\,\pi)^{3/2}\,m\,A\,
{{\Gamma (n-1/2)}\over{\Gamma (n+1)}}\,\Psi^n (r) \,\,. 
\end{eqnarray}

To calculate the relative potential $\Psi (r)$, we use Poisson's equation. For
our spherical system, this is
\begin{equation}
{1\over r^2}\,{{d{\phantom r}}\over{dr}}\!
\left(r^2\,{{d\Psi}\over{dr}}\right) \,=\, -4\,\pi\,G\,\rho_0
\left({\Psi\over\Psi_0}\right)^n \,\,,
\end{equation}
where $\rho_0$ and $\Psi_0$ are the central values of $\rho (r)$ and 
$\Psi (r)$, respectively. We define a dimensionless potential as 
\hbox{$\psi\,\equiv\,\Psi/\Psi_0$}, and a dimensionless radius as 
\hbox{$\xi\,\equiv\,r/r_0$}, where the scale radius $r_0$ is 
\begin{equation}
r_0 \,\equiv\,\sqrt{\Psi_0\over{4\,\pi\,G\,\rho_0}} \,\,.
\end{equation}
Since \hbox{$\rho\,=\,\rho_o\,\psi^n$}, the new potential obeys the Lane-Emden 
equation:
\begin{equation}
{1\over\xi^2}\,{{d{\phantom r}}\over{d\xi}}
\left(\xi^2\,{{d\psi}\over{d\xi}}\right) \,=\, -\psi^n \,\,,
\end{equation}
with boundary conditions \hbox{$\psi (0)\,=\,1$} and 
\hbox{$\psi^\prime (0)\,=\, 0$}. The
nondimensional tidal radius \hbox{$\xi_t\,\equiv\,r_t/r_0$} is the point where
$\psi$ falls to zero.\footnote{The value of $\xi_t$ is derived within each 
polytropic model. We stress that, despite the nomenclature, this ``tidal" 
radius bears no relation to the truncation created by the Galactic potential. 
In Section 3.3, we describe simulations that include the external field.}

For any chosen polytropic index $n$, equation~(10) can readily be solved
numerically. Our task is to translate this nondimensional solution into a
physical model of the initial cluster. Given $n$, the basic quantities 
characterizing the cluster are: $N_{\rm tot}$, the total number of stars; \
$r_v$, the virial radius; and $m$, the mean stellar mass.\footnote{Unlike 
$N_{\rm tot}$ and $r_v$, the mean mass $m$ is not an independent parameter, but
follows from our specified mass function and prescription for binaries; see
\S 2.1.2 below.} The virial radius is defined by
\begin{equation}
r_v \,\equiv\, -{{G\,M^2}\over{2\,W}} \,\,.
\end{equation}   
Here, \hbox{$M\,\equiv\,N_{\rm tot}\,m$} is the total cluster mass, while
$W$ is the gravitational potential energy:
\begin{equation}
W \,=\,{1\over 2}\,\int_0^{r_t}\!4\,\pi\,r^2\,\rho\,\Phi\,dr \,\,. 
\end{equation}
In Appendix A, we show how to obtain the dimensional quantities $r_0$, 
$\rho_0$, and $\Psi_0$ from our three input parameters and the solution 
$\psi (\xi)$.

Given the scale factors $r_0$ and $\rho_0$, we know the dimensional mass
density $\rho (r)$. We then populate space with stars according to a 
normalized distribution $p_1 (r)$ such that \hbox{$p_1 (r)\,dr$} is the
probability a star is located between $r$ and \hbox{$r\,+\,dr$}. This
probability is simply
\begin{equation}
p_1 (r) \,=\, {{4\,\pi\,r^2\,\rho}\over{N_{\rm tot}\,m}} \,\,.
\end{equation}
The actual position vector of each star is then distributed isotropically 
within each radial shell. 

Finally, we require an analogous distribution for stellar speeds. At a given 
radius, the speed must be consistent with the prescribed energy distribution. 
Let $p_2(v|r)\,dv$ be the probability that the speed lies between $v$ and 
$v\,+\,dv$ {\it given} that its radius is $r$. Clearly,
\begin{equation}
p_2 (v|r)\,dv \,\times\, p_1(r)\,dr \,=\, {{f\,d^3v\,d^3r} \over N_{\rm tot}} 
\,\,. 
\end{equation}
Replacing $d^3 r$ by $4\,\pi\,r^2\,dr$ and $d^3 v$ by $4\,\pi\,v^2\,dv$, we 
have, after using equation~(13),
\begin{equation}
p_2 (v|r) \,=\, {{4\,\pi\,v^2\,m\,f}\over\rho}  \,\,.
 \end{equation} 
We take \hbox{$f\,=\, f ({\cal E})$} from equation~(1) and use the definition
of $\cal E$ from equation~(2), finding
\begin{equation}
p_2 (v|r) \,=\, {2\over\pi} \,{{\Gamma (n+1)}\over{\Gamma (n-1/2)}}\,
\Psi^{-3/2}\, \left[ 1\,-\, {v^2\over {2\,\Psi}}\right]^{n-3/2} v^2
\,\,.
\end{equation}
Here, the relative potential is calculated at each $r$ from
\hbox{$\Psi\,=\,\Psi_0\,\psi(\xi)$}, where we recall that $\xi$ is the
nondimensional radius. Given the stellar speed, i.e., the magnitude of the 
velocity vector, the direction of that vector is again distributed 
isotropically in space.

\subsubsection{Stellar Masses: Single and Binary}

Thus far, we have described a cluster that is composed of members with 
identical mass. In actual practice, we assign masses to the stars according to 
a realistic distribution. The parameters of this mass function for the initial
cluster are among those we vary to obtain an optimal match between the evolved
system and the present-day Pleiades. In the course of evolution, some stars
will be given enough energy, through three-body interactions, to escape the
cluster. The most massive ones die out over 125~Myr. It is therefore not 
obvious that the initial mass function is identical to that found today.

We suppose that the distribution of stellar masses in the young Pleiades was
similar in form to the initial mass function for the field population. In 
recent years, large-scale surveys of low-luminosity objects, combined with 
spectroscopy, have established an accurate initial mass function down to the 
brown dwarf limit \citep[e.g.,][]{c08}. The consensus is that the original 
power law of \citet{s55} for masses above solar is joined at the lower end by 
a lognormal function. This basic form appears to hold in diverse environments, 
including young clusters \citep{c05}.

Let $\phi (m)\,dm$ be the probability that a star's mass (in solar units)
is between $m$ and $m+dm$. We posit that this probability is 
\begin{equation}
\phi (m) = \left\{
\begin{array}{rl}
  ({B/m})\,\,{\rm exp}\,\left({-y^2}\right) & \,\, m_{\rm min}\,\le\,m\le \mu\\
  C\,m^\alpha                     & \,\,\mu\,\le\,m\,\le\,m_{\rm max}\,\,,
\end{array} \right.
\end{equation}
where $B$, $C$, $\alpha$, and the joining mass $\mu$ are all constants. We set
\hbox{$m_{\rm min}\,=\,0.08$} and \hbox{$m_{\rm max}\,=\,10$} (although we also
tested a higher mass limit; see \S 3.2 below). The variable
$y$ is given by  
\begin{equation}
y\,\equiv\,{{{\rm log}\,m \,-\, {\rm log}\,m_0}\over{{\sqrt{2}}\,\,\sigma_m}} 
\,\,.
\end{equation}
Here, $m_0$ is the centroid of the lognormal function, and $\sigma_m$ its
width. For input parameters $\alpha$, $m_0$, and $\sigma_m$, the constants $B$,
$C$, and $\mu$ are determined by the normalization condition
\begin{equation}
\int_{m_{\rm min}}^{m_{\rm max}}\!\phi (m)\,\,dm \,=\, 1 \,\,,
\end{equation}
and by requiring that $\phi (m)$ and its first derivative be continuous at
\hbox{$m\,=\,\mu$}. Analytic expressions may be found for the three constants,
which we do not display here.

Most stars are not single objects, but have binary companions. Indeed, the
Pleiades today is especially rich in binaries (see \S 3.4 of Paper~I). Such
pairing must have been present in the initial cluster. We therefore view the
positions and velocities established in the previous subsection as pertaining 
to $N_{\rm tot}$ {\it stellar systems}, rather than individual stars. 
Similarly, the symbol $m$ used, e.g., in equation~(13), actually denotes the
{\it average system mass}, after accounting for binaries. We specify the global
binary fraction as a parameter $b$, which gives the probability that a system 
actually consists of two stars. Conversely, a fraction \hbox{$1-b$} of the 
systems are indeed single stars. Their mass is distributed according to the 
probability $\phi (m)$.

Our analysis of the present Pleiades in Paper~I showed that the masses of the
component stars within binaries are correlated. Such a correlation must also
have been present at early times. Accordingly, we include the effect in our 
initial state. Within the fraction $b$ of systems that are binaries, we first 
independently assign masses to each component, using the probability 
distribution $\phi (m)$. After identifying the primary mass $m_p$ and the
secondary mass $m_s$, we then alter the latter to $m_s^\prime$, where
\begin{equation}
m_s^\prime\,=\,m_s \left({m_p\over m_s}\right)^\gamma \,\,. 
\end{equation}
Here, $\gamma$ is an input parameter that measures the degree of mass 
correlation within binaries (see also eq.~(42) of Paper~I). If 
\hbox{$\gamma\,=\,0$}, the component masses are uncorrelated, while 
\hbox{$\gamma\,=\,1$} corresponds to perfect matching. 

We give our binaries randomly inclined orbital planes, and a period and 
eccentricity distribution characteristic of present, solar-type binaries both 
in the field \citep{dm91} and in the Pleiades itself \citep{b97}. If $p_3 
({\cal P})\,d{\cal P}$ is the fraction of systems with periods between $\cal P$
and ${\cal P} + d{\cal P}$, then $p_3 ({\cal P})$ is lognormal:  
\begin{equation}
p_3 ({\cal P})\,=\,{1\over{\sqrt{2\pi} \,\sigma_{\cal P}}}\,
{\rm exp}\,\left(-z^2\right) \,\,,
\end{equation}
where
\begin{equation}
z \,\equiv\, {{{\rm log}\,{\cal P} \,-\, {\rm log}\,{{\cal P}_0}}\over
\sqrt{2}\,\sigma_{\cal P}} \,\,.
\end{equation}
We set the centroid period to \hbox{${\rm log}\,{\cal P}_0 \,=\, 4.8$} and the
width to \hbox{$\sigma_{\cal P} \,=\, 2.3$}, where the period is measured in 
days. The eccentricity distribution has a thermal distribution:
\begin{equation}
p_4 (e) \,=\, 2\,e, 
\end{equation}
as motivated both by observations \citep{dm91} and theory \citep{h75}.

The initial cluster described thus far is homogenous, in the sense that any 
volume containing an appreciable number of systems has the same average system 
mass. However, there have long been claims of observed {\it mass segregation} 
in young  clusters, i.e., an increase of average stellar mass toward the 
center \citep{s88,js91,m97,s06}. The present-day Pleiades also exhibits this
phenomenon to a striking degree (see \S 4.2 of Paper~I). We want to see if this
property developed on its own or was inherited from an earlier epoch. We
accordingly include a quantitative prescription for mass segregation in our 
initial state.

One system has a higher probability of being near the cluster center than 
another, in a time-averaged sense, if its relative energy $\cal E$ is greater. 
Mass segregation therefore manifests itself as a correlation between the
system mass $m$ and $\cal E$. This fact was noted by \citet{b08}, who used it
to implement a specific procedure for mass segregation. Here we have adopted 
a variant of their method that allows us to include the effect to a variable 
degree. We first assign $\cal E$- and $m$-values to all member systems 
according to equation~(1) and our prescription for binary masses. We 
then place the systems in two lists - the first ordered by increasing $m$, and 
the second by increasing $\cal E$. When we first construct these lists, the 
ranking of the system in the first is unrelated to its ranking in the second. 
This is the case of zero mass segregation. There would be perfect mass 
segregation if the two rankings were identical.
 
Let us quantify the intermediate case. For a star of given mass, we find its
index in the mass-ordered list. To assign an energy to that star, we choose the
second (energy-ordered) index from a Gaussian distribution centered on the 
mass index. The width of this distribution, denoted $\sigma_{\cal E}$, can 
be infinite (no mass segregation) or zero (perfect segregation). More 
generally, we define a parameter $\beta$, the degree of mass segregation, which
varies between 0 and 1. After some trial and error, we adopted the following 
prescription relating the width $\sigma_{\cal E}$ to $\beta$:
\begin{equation}
\sigma_{\cal E} \,=\, -{1\over 2}\,N_{\rm tot}\,\,{\rm ln}\,\beta \,\,.
\end{equation}
The logarithmic dependence on $\beta$ ensures that $\sigma_{\cal E}$ has the
desired behavior in the extreme limits. The proportionality with $N_{\rm tot}$
ensures that our algorithm gives the same degree of biasing in clusters of any
population.

In summary, $\beta$ becomes another input parameter that we vary within the
initial configuration. As we will see, having a non-zero $\beta$ is critical
to obtaining a proper match between the evolved cluster and the Pleiades today.
Mass segregation was therefore present at a relatively early epoch. 

Since relatively massive stars preferentially reside near the center, our
imposition of mass segregation alters the shape of the gravitational potential 
from that of a single-mass polytrope. Relative to the total gravitational 
energy, the total kinetic energy has a value slightly below that for virial 
equilibrium. We rescaled all stellar velocities by a uniform factor to restore 
exact equilibrium. In practice, this factor was typically about 1.05. 

For convenient reference, Table~1 lists the full set of our input parameters 
for the starting state. Anticipating the results detailed in \S 3 below, the 
last column gives the numerical value of each parameter in the optimal 
configuration. We also list the associated uncertainties. The meaning of these
uncertainties and how they were assessed, will also be discussed presently.

\subsection{Characterizing the Evolved Cluster}

After evolving a particular initial state for 125~Myr, we compare the outcome
with the actual Pleiades. In making this comparison, it is important to 
``observe'' the simulated cluster under the same conditions as the real one.
Thus, we project the three-dimensional distribution of stars onto a 
two-dimensional plane, assumed to lie at the mean Pleiades distance of
133~pc \citep{s05}. The angular separation $\Delta\theta$ between each pair of
stars is then determined. If $\Delta\theta_{\rm res}$ denotes the telescope
resolution, then any pair with \hbox{$\Delta\theta\,<\,\Delta\theta_{\rm res}$}
is taken to be an unresolved point source. For the near-infrared catalog of
the Pleiades analyzed in Paper~I \citep{s07}, an appropriate value of
$\Delta\theta_{\rm res}$ is 10~arcsec. Note that our unresolved sources
include a small fraction (less than 0.5~percent) of triples and high-order
systems, as well as a few unrelated pairs observed to be close in projection.
We denote as $N_s$ the total number of point sources out to a radius from the
cluster center of 12.3~pc, corresponding in angle to $5.3^\circ$. This was the 
radius enclosing the catalog of sources used in Paper~1. For each 
simulated evolution, we compare the final $N_s$-value with the observed 
Pleiades figure.

The vast majority of stars observed in the Pleiades today are on the main
sequence (see Fig.~1 of Paper~1). The number of post-main-sequence objects,
while relatively small, is sensitive to the shape of the stellar mass
function. Hence, it is important that we reproduce, as closely as possible,
the number inferred for the present-day cluster. At an age of 125~Myr, the
main-sequence turnoff is about $4\,\,\Msun$. If $N_4$ denotes the number
of stars (singles or primary stars in binaries) whose mass exceeds 
$4\,\,\Msun$, then this quantity, as calculated, may also be compared directly
with that in the Pleiades. Similarly, we compare $M_{\rm tot}$, the total mass
of all stars in the evolved cluster, with the Pleiades mass obtained through
the statistical analysis of Paper~1. 

One striking result of Paper~1 was the prevalence of binaries. Specifically,
we found that the near-infrared fluxes of the catalogued point sources 
demanded that the fraction \hbox{$b_{\rm unres}\,=\,0.68$} were unresolved 
binaries. (Any resolved binaries were listed in the catalog as separate 
sources.) For our assumed resolution limit $\Delta\theta_{\rm res}$, we could 
also assess $b_{\rm unres}$ computationally for each evolutionary run. Again,
this is the fraction of point sources representing two or more unresolved 
stars. Note that $b_{\rm unres}$ is less than the {\it initially} imposed
binary fraction $b$, both because some pairs are wide enough to be resolved,
and because others are torn apart in the course of evolution.

The distribution of stellar masses is, of course, another property that should
be compared with the actual Pleiades. As just described, we set the form of the
distribution within the initial configuration as a lognormal function with a
power-law tail. The apportionment of masses within the evolved state could in
principle differ, due to the escape of some stars from the cluster and the
death of others with sufficiently high mass. Our procedure is first to find,
within the output state, the normalized distribution of single stars. Included
in this distribution are both isolated stars and the components of 
{\it resolved} binaries. We then peer within {\it unresolved} binaries and find
the analogous distributions of primary mass $m_p$ and secondary mass $m_s$. 
Finally, we record the distribution of the binary mass ratio, 
\hbox{$q\,\equiv\,m_s/m_p$}.

In Paper I, we statistically determined the stellar masses from the photometric
data by assuming that the single-star distribution was a pure lognormal, with
centroid $m_0$ and width $\sigma_m$. For consistency, we characterize the 
evolved cluster in our simulations in a similar fashion. Now for a given 
single-star function and binary correlation parameter $\gamma$, the primary,
secondary, and $q$-distributions are all uniquely determined. Appendix~B 
outlines the mathematical derivation. The task is to vary $\gamma$, as well as
$m_0$ and $\sigma_m$, for the presumed lognormal single-star distribution
until this function, as well as the primary, secondary, and $q$-distributions,
best fit those we find directly in the numerical output. We then compare
$\gamma$, $m_0$, and $\sigma_m$ to these same quantities derived in a
similar way for the observed Pleiades.   

We next consider the projected density profile. We divide the cluster into
radial bins that match those used in the analysis of the Pleiades. The 
resulting surface density of stellar systems is then fit to the empirical
prescription of \citet{k62}:
\begin{equation}
\Sigma (R) \,=\, k \left({1\over\sqrt{1 + (R/R_c)^2}} 
\,-\, {1\over\sqrt{1 + (R_t/R_c)^2}}\right)^2 \,\,.
\end{equation}
Here, $R$ is the projected radius, $k$ is a constant with the dimensions of a
surface density, and $R_c$ and $R_t$ are the core and tidal radii, 
respectively. We determine the values of $k$, $R_c$ and $R_t$ which best match 
the data, i.e., the same parameters determined for the real Pleiades by an 
analogous fitting procedure. However, only $R_c$ is used in the final 
optimization routine (see below). We also determine the King concentration 
parameter \hbox{$c_K \,\equiv\,{\rm log}\,(R_t/R_c)$}, both for
each evolved simulation and in the real cluster. From equation~(25), the 
central surface density $\Sigma_0$ is  
\begin{equation}
\Sigma_0 \,=\, k \left(1\,-\,  {1\over\sqrt{1 + (R_t/R_c)^2}}\right)^2 \,\,.
\end{equation}
This is also compared to the Pleiades value.

Finally, we measure the degree of mass segregation. For the evolved cluster, we
compute the cumulative fraction of systems contained within a projected radius
$R$, both by number ($f_N (R)$) and mass ($f_M (R)$). As in Paper ~I (\S 4.2),
the Gini coefficient is computed as
\begin{equation}
G \,=\, 2 \int_0^1\!\left(f_M - f_N\right)\,df_N \,\,.
\end{equation}
and then compared to that found in the Pleiades.
 
Table~2 gives the full list of quantities evaluated for each evolved cluster.
We also display the values found when the optimal initial state is used, as 
well as the corresponding figures in the actual Pleiades. Notice that
$m_0$ and $\sigma_m$ for the mass function do not match those in the initial
state, as given in Table~1. This discrepancy arises partly from real changes
of the stellar masses, but even more from our adoption of a simple lognormal 
when fitting the evolved cluster. The tabulated errors for the calculated
quantities were obtained by running 25 simulations, all with identical input 
parameters. Since we populated the cluster stochastically, according to 
probability distributions (e.g., $\phi (m)$ in eq.~(17)), initial states 
differed from one another in detail. The errors represent the standard
deviations for each quantity in the evolved cluster, due solely to differing
realizations of the initial state. The tabulated errors for the observed
Pleiades are from the calculation of Paper~I. In addition, $N_s$ and $N_4$ are
assumed to be Poisson-distributed, so that the errors are $\sqrt{N_s}$ and
$\sqrt{N_4}$, respectively.   

\subsection{Optimization Procedure}

As a first guess, we set all the input parameters equal to values appropriate
for the Pleiades today. In Paper~I, we characterized the present-day cluster
as a King model, so we initially adopted this prescription. The analytic
models of \citet{k66} have the distribution function 
\begin{equation}
f ({\cal E}) = \left\{
\begin{array}{rl}
  A\,{\rm exp}\,
\left({W_0\,{\cal E}}/\Psi_0\right)\,-\,1
  & \,\,{\cal E}\,\ge\,0\\
  0                   & \,\,{\cal E}\,<\,0\,\,.
\end{array} \right.
\end{equation}
Here $\Psi_0$, the central value of the relative potential, is again set by our
basic input quantities. The dimensionless parameter $W_0$, like $c_K$, 
characterizes the degree of central concentration.\footnote{The relation
of $W_0$ to $c_K$ is shown in Figure~(4-10) of \citet{bt87}. The independent 
variable in their plot, called $\Psi_0/\sigma^2$, is precisely $W_0$.} 
Running a King model for 125~Myr, we found that it invariably became more
centrally concentrated than the actual Pleiades. We therefore tried 
successively lower $W_0$-values for the initial state. Now equation~(28) shows 
that $f({\cal E})$ is proportional to $\cal E$ in the limit of small
$W_0$. Comparison with equation~(1) reveals that such a model is
equivalent to a polytrope of \hbox{$n\,=\,5/2$}. Our search for 
low-concentration initial states therefore led us naturally to the polytropic 
models described in \S 2.1. Within the regime of polytropes with small $n$, we 
varied other input parameters, such as $r_v$, as necessary.

Once our computed cluster began to resemble the Pleiades, we changed to a more
systematic gradient method for refining the initial state. Let $\bx$ be the
vector whose elements are the 9 input parameters listed in Table~1. Similarly,
let $\by$ represent the 11 evolved cluster properties of Table~2. This latter
vector is, of course, a function of $\bx$. To move $\by$ toward the values
characterizing today's Pleiades, we need to evaluate, in some sense, the
gradient of this function.

A practical complication is one to which we alluded earlier. Even among
evolutionary runs assuming an identical input vector $\bx$, the resulting $\by$
differs because of the stochastic sampling of the various assumed distribution
functions. In computing the gradient, we need to take a step size $\bh$
large enough that the resulting change in $\by$ exceeds that due to this
realization variance. We found that the prescription
\hbox{$\bh \,=\, 0.5\,\bx$} sufficed for this purpose \citep[see \S 5.7 of] 
[for a more rigorous justification]{p02}.

For each $\bx$, we first do 9 runs and average the result to obtain
\hbox{$\by (\bx)$}. We then decrease, in turn, each element $x_j$ to
\hbox{$x_j \,-\,h_j$}, and find the average output of two runs at
each decreased $x_j$-value. Similarly, we find the average result of two runs
at each \hbox{$x_j\,+\,h_j$}. We thus establish the \hbox{$11 \times 9$}
matrix of derivatives $\cal D$, whose elements are 
\begin{equation}
{\cal D}_{ij} \,\equiv \, {{y_i (x_j + h_j) \,-\,
y_i (x_j - h_j)}\over {2\,h_j}} \,\,. 
\end{equation}   
The change in outputs for any subsequent input change $\Delta x$ may then be
approximated by
\begin{equation}
\Delta\by \,=\, {\cal D}\,\Delta\bx \,\,.
\end{equation}
Here, the vector $\Delta\by$ is taken to be the difference between the current
$\by$-vector and that for the Pleiades. We may evaluate the 9 elements
$\Delta x_j$ by solving the 11 linear equations summarized in~(29). Since the
system is overdetermined, we did a least-squares fit to find that set of
$\Delta x_j$ which best satisfied the equations. 

As we took a step in $\bx$, we evaluated how close the resulting $\by$ was to
${\by}_p$, the aggregate properties of the observed Pleiades. We did a 
$\chi^2$-test, where
\begin{equation}
\chi^2 \,=\, \sum_{i=1}^{11} {{\left(<y_i>\,-\,y_{p,i}\right)^2}\over
\sigma_i^2}\,\,.
\end{equation}
Here, each $<\!y_i\!>$ is the average $y_i$ value, established by doing 9 runs
with identical input values. The standard deviation $\sigma_i$ includes
errors in both the inferred Pleiades properties and those generated by 
different statistical realizations of the input state:
\begin{equation}
\sigma_i^2 \,\equiv\, \sigma_{p,i}^2 \,+\, \sigma_{<y_i>}^2 \,\,.
\end{equation}   
The first righthand term is the Pleiades variance whose square root is the
error given in the last column of Table~2. The quantity $\sigma_{<y_i>}^2$ is 
the error in the {\it mean} $y_i$. This error in the mean is related to 
$\sigma_{y,i}$, the variance in each individual $y_i$, by
\begin{equation}
\sigma_{<y_i>}^2 \,\equiv\, {1\over 9}\,{\sigma_{y,i}^2}\,\,.
\end{equation}

For the first few $\bx$-steps, $\chi^2$ declined, but then stalled. Beyond this
point, the gradient method itself was clearly failing, as it indicated initial
states which evolved to configurations {\it less} resembling the Pleiades. The
difficulty was that the numerical derivatives of equation~(29) were too crude
to refine the initial state further. Refinements are possible in principle, but
prohibitive computationally. After pushing the method to its limit, we were
forced to stop the search before $\chi^2$ reached a true minimum. We took the 
last state in the sequence where $\chi^2$ declined to be the best-fit initial 
configuration. 

Our final task was to assess the errors in all input parameters for this state.
These should reflect uncertainties in properties of the actual Pleiades, 
as well as the variation in output parameters among different runs using 
identical inputs. This latter effect is quantified by the covariance matrix
$\cal Y$, whose elements are
\begin{equation}
{\cal Y}_{ij} \,\equiv\, 
\left<\left(y_i-<\!y_i\!>\right)\left(y_j-<y_j>\right)\right> \,\,.
\end{equation}
The averaging here refers to different realizations using identical input
parameters. Standard error propagation \citep[][Section 1.6]{c98} dictates that
the known $\cal Y$ is related to $\cal X$, the desired covariance matrix of 
input parameters, through the derivative matrix and its transpose:
\begin{equation}
{\cal Y} \,=\, {\cal D}\,{\cal X}\,{\cal D}^T \,\,.
\end{equation}
We need to invert this equation to obtain $\cal X$. As noted, the input errors
should also reflect the observational uncertainties in the Pleiades itself. We
do not know the correlation of these observational uncertainties. Thus, we use
on the lefthand side of equation~(35) a matrix ${\cal Y}^\prime$, formed by
adding $\sigma_{p,i}^2$ to each diagonal element ${\cal Y}_{ii}$.

Since $\cal D$ is not a square matrix, a standard inverse cannot be defined.
However, the product \hbox{${\cal D}^T {\cal D}$} {\it is} square, and so has 
an inverse, provided it is not singular. As discussed in \citet{g83}, this fact
allows us to define the pseudo-inverse of $\cal D$:
\begin{equation}
{\cal D}^+ \,\equiv\, \left({\cal D}^T {\cal D}\right)^{-1}\,{\cal D}^T \,\,.
\end{equation}
The term ``pseudo-inverse'' is appropriate since
\begin{equation}
{\cal D}^+ \, {\cal D} \,=\,  
\left({\cal D}^T {\cal D}\right)^{-1}\,{\cal D}^T\,{\cal D} \,=\, {\cal I}\,\,,
\end{equation}
where $\cal I$ is the identity matrix. Taking the transpose of this last
equation, we also find
\begin{equation}
\left({\cal D}^+\,{\cal D}\right)^T \,=\,
{\cal D}^T\,\left({\cal D^+}\right)^T \,=\, {\cal I}^T \,=\, {\cal I} \,\,.
\end{equation}
By employing equations~(37) and (38), inversion of the modified equation~(35)  
is straightforward:
\begin{eqnarray}
{\cal D}^+\,{{\cal Y}^\prime}\,\left({\cal D}^+\right)^T \,&=&\,
{\cal D}^+\,{\cal D}\,{\cal X}\,{\cal D}^T\,\left({\cal D}^+\right)^T 
\nonumber \\
&=&\, {\cal X} \,\,.
\end{eqnarray}
The errors in the initial cluster parameters of Table~1 are then the standard
deviations obtained from the diagonal elements of $\cal X$.

\section{Numerical Results}

\subsection{Global Properties of the Cluster}

Table 1 lists the optimal values for the parameters characterizing the initial
state. The polytropic index $n$ is about 3, corresponding to a volumetric,
center-to-average, number density contrast of 54.\footnote{Because of mass
segregation, the center-to-average contrast in the volumetric mass density is
higher, about 100.} This particular polytrope closely resembles a \citet{k66} 
model with \hbox{$W_0 \approx\,1.4$}. Note the relatively large uncertainty in 
the optimal $n$, reflecting the fact that a range in initial density contrasts 
relaxes to a similar state after 125~Myr. There is much less uncertainty in the
virial radius $r_v$, which is surprisingly large compared to observed embedded 
clusters (see \S 4 below). Smaller assumed $r_v$-values, however, evolved to 
systems with too high a density contrast.

Figure~1 shows, as the dashed curve, the initial surface density as a function 
of projected radius. Also plotted ({\it solid curve}) is the evolved surface 
density, along with observed data from the Pleiades. Notice how the surface 
density {\it decreases} with time, a result of the inflation experienced by the
entire cluster. This behavior contrasts with expectations from the standard 
acount of dynamical relaxation \citep[e.g.][Chapter 8]{bt87}. The swelling of 
the central region that we find is consistent, however, with previous 
simulations of {\it binary-rich} clusters with relatively low populations 
\citep{pz01}. We explore further the underlying physical mechanism in \S 3.3 
below.

Note from Table~1 that $N_{\rm tot}$, the initial number of stellar systems, is
determined to within about 5~percent uncertainty. The main constraint here is 
the need to match $N_s$, the final, observed number of point sources. Note also
that \hbox{$N_{\rm tot} \,<\, N_s$} throughout the evolution. Almost all the 
stellar systems are binaries. Some of these are wide enough that they could be 
resolved observationally. Thus, the total number of point-like (i.e., 
unresolved) sources is always higher than $N_{\rm tot}$, the number of stellar 
systems (resolved or unresolved). By the same token, the unresolved binary
fraction, \hbox{$b_{\rm unres}\,=\,0.68$}, is significantly less than the
full initial binary fraction, \hbox{$b\,=\,0.95$}. Indeed, we were forced to
choose a $b$-value close to unity in order to make $b_{\rm unres}$ close to 
the observationally inferred figure (see Table~2).

Figure~2 quantifies the degree of mass segregation in the evolved cluster, 
Following the technique introduced in Paper~I, we plot $f_M$, the fractional
cumulative mass at any projected radius, against $f_N$, the fractional 
cumulative number. The fact that this curve rises above the dashed diagonal 
(\hbox{$f_M \,=\,f_N$}) indicates the existence of mass segregation. The 
empirical \hbox{$f_M - f_N$} relation for the Pleiades, shown by the points 
with error bars, is well matched by the simulation. We were able to obtain this
match only by adopting a non-zero value of $\beta$, the {\it initial} degree of
mass segregation defined in equation~(24).\footnote{Table~1 lists, for
convenience, a symmetrical error on the best-fit $\beta$. Although the lower
bound is accurate, even higher values give acceptable results, due to the 
saturation of mass segregation described in \S 3.2 below.}

Four quantities in Table~1, $m_0$, $\sigma_m$, $\alpha$, and $\gamma$, concern
the mass function. The number of stars escaping the cluster during its 
evolution is relatively small (on average, 280 of the 2400 stars present
initially). Because of this small loss, and because few members evolve off the
main sequence, the initial and final mass functions are essentially 
identical, and all four parameters are highly constrained by the 
observations. Note, in particular, that the exponent $\alpha$ of the power-law 
tail directly influences $N_4$, the observed number of massive stars. The 
binary correlation parameter $\gamma$ is independent of the single-star mass 
function, but influences the primary, secondary, and $q$-distributions, as
described in \S 2.2. Any substantial variation in $\gamma$ would alter the 
corresponding parameter obtained statistically for the observed cluster (see 
Fig.~5 of Paper I).

Figure~3 compares our {\it evolved} single-star mass function with the 
Pleiades. The solid curve is a lognormal fit to the simulation result, which
is fully characterized by $m_0$ and $\sigma_m$ in Table~2. The data points,
along with error bars, represent the inferred single-star mass function for the
Pleiades, obtained through the maximum likelihood analysis of Paper~I. The
agreement with the simulation is naturally poorest at the highest masses, since
we modeled the output as a pure lognormal, in order to be consistent with the
procedure adopted in Paper~I.

Finally, Figure~4 shows the initial distribution of the binary mass ratio $q$.
We see how nearly equal-mass systems are strongly favored for our best-fit 
$\gamma$ of 0.73. In the simulations, this distribution evolves almost 
unchanged, and closely matches the one inferred for the Pleiades today.
The figure also displays the $q$-distribution for the hypothetical case of
\hbox{$\gamma\,=\,0$}. Such random pairing of stellar masses does not result
in a flat curve, as one might expect. Instead, it reflects the character of the
single-star mass function, which here is lognormal. As seen in Figure~4, the
$q$-distribution for \hbox{$\gamma\,=\,0$} peaks at \hbox{$q\,=\,0.34$} and 
still vanishes as $q$ approaches 0.   

\subsection{Past Evolution}

We can now describe, based on our suite of simulations, the evolution of the
cluster from its initial state to the present epoch. The main trend is an
overall expansion of the system. This tendency is clear in Figure~5, which
shows the variation in time of the virial radius, $r_v$. After an initial
drop, lasting about two crossing times 
\hbox{$(t_{\rm cross}\,=\,10\,\,{\rm Myr})$} the radius steadily swells, 
increasing by about 40~percent to the present. From the definition of $r_v$ in
equation~(11), we infer that the gravitational potential energy $W$ is
decreasing in absolute magnitude, i.e., the cluster is gradually becoming
less bound. Note that we do not obtain $r_v$ by calculating
$W$ directly, but through fitting the cluster at each time to a King model,
and then finding the appropriate $r_v$ for the best-fit model parameters. 

Figure~5 shows that $R_c$, the projected core radius, displays similar behavior
to $r_v$. After the transient phase which again lasts about two crossing times,
$R_c$ also swells, albeit more slowly. Analogous early adjustments are evident 
in other global quantities (see Figs. 6 - 8). This transient results from our 
implementation of mass segregation, which alters slightly the gravitational
potential (recall Section 2.1.2). Although the initial cluster is in virial
equilibrium, the stellar distribution function is no longer a steady-state 
solution to the collsionless Boltzmann equation. Within the first two crossing 
times, the distribution readjusts to become such a solution. The core radius 
bounces, before settling to a value that subsequently evolves more gradually.

Expansion of a cluster's outer halo is one manifestation of dynamical 
relaxation. However, application of equation~(4-9) of \citet{bt87}, with
\hbox{${\rm ln}\,\Lambda\,=\,{\rm ln}\,(0.4\,N)$}, reveals that the relaxation 
time is 250~Myr, or about twice the Pleiades age. In addition,
the inner cores of relaxing systems shrink, giving energy to the halo. The
secular expansion of $R_c$ further indicates that we are {\it not} witnessing
the usual effects of dynamical relaxation. Figure~6 provides yet another
illustration of this point. Here, we see that the King concentration
parameter $c_K$ remains virtually constant, again following an initial 
adjustment. Recall that \hbox{$c_K \,\equiv\,{\rm log}\,(R_t/R_c)$}, where 
$R_t$ is the projected tidal radius. Thus, $R_t$ and $R_c$ swell at about the 
same pace. 

The projected surface number density, $\Sigma (R)$, currently peaks strongly at
\hbox{$R\,=\,0$} (Fig.~1). The actually central value, however, previously 
declined from an even higher level. Figure~7 shows this gradual decline, which
is consistent with the previously noted rise in $R_c$. Thus, $R_c$ increases
from 1.6 to 2.2~pc over the period from \hbox{$t\,=\,30\,\,{\rm Myr}$} to
\hbox{$t\,=\,125\,\,{\rm Myr}$}. Over the same interval, $\Sigma_0$ falls by
a factor of 0.50, which is close to $(1.6/2.2)^2$. The number of systems in 
the core therefore remains virtually constant as the core itself expands. The 
volumetric number density similarly falls in the central region.

In Paper~I, we documented a strong degree of mass segregation in the current
Pleiades, quantifying this property through the Gini coefficient. Another
result of the current study is that $G$ did {\it not} attain its current value
through purely stellar dynamical evolution. As seen in the top curve of 
Figure~8, $G(t)$ rose only slightly at first, and then remained nearly 
constant, even declining somewhat in the recent past. Initial states in which 
the parameter $\beta$ was too low never attained the requisite degree of mass 
segregation. As an illustration, Figure~8 shows also the result from a single 
simulation using $\beta\,=\,0$ initially. The Gini coefficient does grow, but 
not by enough to match observations. We note, parenthetically, that $G(t)$ 
exhibits oscillatory behavior over the a period that roughly matches the 
crossing time. These oscillations (unlike the initial readjustment)  were 
washed out in the averaging procedure that produced the top curve in the 
figure. Finally, we remark that $G(t)$ appears to saturate in time. We will 
return to this interesting phenomenon shortly.

All the simulations we have described thus far ignored any effects of stellar
evolution. We could afford this simplification because of the relatively small
number of cluster members that would have evolved significantly over 125~Myr.
However, the code Starlab does have the capability of tracking stellar 
evolution, including mass loss, through fitting formulae. As a check, we
retained our usual maximum mass of $10\,\,\Msun$ and ran 25 simulations using 
the best-fit initial cluster parameters, but with stellar evolution included. 
The mass loss from relatively massive cluster members did not have a 
significant dynamical effect, and the endstate of the cluster was essentially
identical. With reference to Table~2, the only parameter that changed 
appreciably was $N_4$, which fell.

In more detail, the few stars above 7~$\Msun$ usually evolved to white dwarfs 
of approximately solar mass. On average, about 10 white dwarfs formed, of
which 7 were the secondaries within binaries. Even the few that were primaries
were faint relative to their main-sequence companions, and thus would be
difficult to detect. Our findings are thus consistent with the observation of 
\citet{f03} that white dwarfs are generally rarer in open clusters than might 
be expected statistically from the initial mass function.

Finally, we relaxed the upper mass limit in the single-star mass function and 
allowed the maximum mass to be arbitrarily large, according to the power law in
equation~(17). Choosing stars stochastically from this distribution yielded a
few members with masses as high as $40\,\,\Msun$. If we again allowed for 
stellar evolution and used our standard initial cluster parameters, the 
evolution {\it did} take a different turn. The very massive stars represented a
significant fraction of the total cluster mass, and their death had a 
quantitative impact. As before, the cluster went through an initial adjustment,
partially from the heavier stellar mass loss. The system then smoothly
expanded, but at a faster pace. At 125~Myr, the projected core radius $R_c$ 
was 3.1~pc, or 1.5 times larger than that of the present-day Pleiades. 
Similarly, the central surface density $\Sigma_0$ was a factor 0.58 lower. Had 
we begun with very massive stars in an initial state a factor of 1.5 
{\it smaller} than our standard one, a closer match would have resulted.

These results were instructive, if somewhat academic. In reality, stars more 
massive than about  $10\,\,\Msun$ would have inflated HII regions so quickly 
as to ionize and disperse the parent molecular cloud forming the Pleiades. 
In order to retain even a remnant, gravitationally bound cluster, the initial
membership must have been very large, about 10,000 stars in the simulations
of \citet{k01}. We stress that even this figure is a lower bound, as
\citet{k01} assumed a star formation efficiency in the parent cloud of
33~percent by mass. Such an efficiency is plausible within individual dense 
cores \citep{a07}, but significantly higher than observational and theoretical 
estimates in cluster-forming clouds \citep[e.g.][]{d82,hs07}.  

Suppose we nevertheless adopt this scenario as a limiting case, and assume 
provisionally that the Pleiades progenitor contained at least 10,000 individual
stars. Such groups are rare. Equation~(39) of \citet{mw97} gives the birthrate 
of OB associations based on their population of supernova progenitors 
\hbox{($m > 8$)}. If we use our adopted initial mass function to estimate this 
population, then the birthrate of relevant OB associations is 
\hbox{$0.09~{\rm Myr}^{-1}~{\rm kpc}^{-2}$}. This is a factor between 5 and 8 
smaller than the total formation rate of open clusters \citep{am01,ms78}. It 
is unlikely, therefore, that formation through dispersing OB associations 
dominates, and we continue to use an upper mass limit of about  $10\,\,\Msun$
for the Pleiades.

\subsection{Future Evolution}

The same calculations that reconstruct the past history of the Pleiades may 
also be used to predict its development far into the future. It is still
believed, following the original proposal by \citet{s58}, that most open 
clusters are eventually destroyed by the tidal gravitational field of passing
interstellar clouds, now identified as giant molecular complexes. In this 
project, we do not attempt to model encounters with such external bodies. 
However, Starlab can follow the effects of the {\it Galactic} tidal field, both
through imposition of the appropriate external potential and by adding a 
Coriolis force to individual systems. We switched on the Galactic field, in 
addition to stellar evolution, and followed the cluster from its initial state 
for a total of 1~Gyr. While most open clusters do not survive this long, some 
do last up to several Gyr \citep{f95}. Our simulation thus models at least a 
portion of the Pleiades' future evolution.\footnote{The very oldest clusters
have large galactocentric radii, and thus experience both a weaker tidal
field and less frequent encounters with giant molecular clouds. Clearly, the
Pleiades does not fall into this category.} 

Up to the present cluster age of 125~Myr, adding the Galactic tidal field
and stellar mass loss made very little difference in the evolution. Beyond this
point, the cluster will continue the overall expansion that characterized it in
the past. As seen in Figure~9, the central density $\Sigma_\circ$ keeps 
declining. The falloff is roughly exponential, with an e-folding time of 
400~Myr. This figure, along with Figures 10 and 11, show average results from 
the 4 runs we conducted. Even after averaging, the calculated $\Sigma_\circ$ 
displays increasing scatter for \hbox{$t\,>\,700\,\,{\rm Myr}$}. By this time, 
the {\it total} population has also fallen to the point that numerical 
determination of $\Sigma_\circ$ (through a fitted King model) becomes 
problematic.

The decline in the cluster population, which was modest until the present, 
accelerates as stars are tidally stripped by the Galactic gravitational field.
It is the lighter stars that populate the cluster's outer halo and that
preferentially escape. Consequently, the average mass of the remaining cluster 
members rises. Figure~10 shows both these trends. Displayed here is $N_s$, the 
number of systems contained within the initial Jacobi radius of 
14.3~pc.\footnote{The Jacobi radius is the spherical average 
of the zero-velocity surface in the presence of the Galactic tidal field 
\citep[][p. 452]{bt87}.} By 1~Gyr, the total membership has fallen to a few 
dozen systems. Meanwhile, the average system mass $\langle m\rangle$ rises, 
almost doubling by the end. Careful inspection of Figure~10 shows that 
$\langle m\rangle$ initially fell slightly to its present-day value. This 
falloff reflects the loss, through stellar evolution, of the most massive 
members, an effect which is eventually overwhelmed by the escape of the 
lightest systems. Notice again the jitter in the $\langle m\rangle$ curve at 
later times.

Despite this qualitative change in the cluster's internal constitution, the
degree of mass segregation remains essentially constant until very late times.
Figure~11 displays the Gini coefficient. In detail, $G(t)$ exhibits 
oscillations qualitatively similar to what we saw in lower portion of Figure~8.
Nevertheless, its average magnitude does not appreciable change until 
\hbox{$t\,\sim\,700\,\,{\rm Myr}$}, when it begins a steep descent. The very 
large scatter during this late epoch again reflects the diminishing population.
  
What accounts for these trends? During most of the evolution, some process is
able to enforce mass segregation, despite the continual depletion of the 
lightest members. This process evidently loses its efficacy at late times, when
the total population falls too low. Earlier, we showed that dynamical
relaxation did {\it not} establish the present-day level of mass segregation.  
Although we are now spanning a period well in excess of the initial relaxation
time (250~Myr), we still do not see the classic behavior - monotonic shrinking
of the core that feeds halo expansion. Up to 
\hbox{$t\,\sim\,200\,\,{\rm Myr}$}, the projected core radius $R_c$ continues
the increase noted earlier. Between 200 and 600~Myr, when 75~percent of cluster
members escape, $R_c$ does decline slightly, from 2.2 to 1.2~pc. Thereafter,
the core swells once more. 

We believe that the system's overall expansion is due principally to the 
release of energy in three-body encounters, specifically, close passages of
binaries and single stars. Over the Gyr time span, mass loss through stellar 
evolution and tidal stripping weakens the cluster's gravitational binding, 
rendering it increasingly responsive to such internal heating. We ascribe
the maintenance of mass segregation, i.e., the inward drift of more massive
stars, to dynamical friction with the background population. We shall revisit
these key processes, binary heating and dynamical friction, momentarily. 

Figure~12 shows graphically how the cluster will appear far in the future. Here
we show positions of the member systems projected into the Galactic plane, both
at the present time and at \hbox{$t\,=\,700\,\,{\rm Myr}$}. One sees at
present a slight elongation along the direction toward the Galactic Center.
This tidal stretching is well documented observationally \citep{rm98}. Stars
that leave tend to do so along that direction. But any appreciable excursion 
leads, because of the Galaxy's differential rotation, to a change in angular 
speed. As a result, two tidal streams develop that are orthogonal to the 
Galactocentric radius. These streams are present in both panels of Figure~12,
but are especially noticeable in the diminished cluster shown at the right.
 
Since we suspected that binaries were important in the gross dynamics of the
cluster, we recalculated the entire 1~Gyr evolution after effectively removing
all primordial binaries from the system. We began with the same, best-fit
initial state as before, which had the usual distribution of single-star masses
and initial binary fraction \hbox{$b\,=\,0.95$}. However, we replaced every 
binary by a single star, located at the system's center of mass and comprising 
the total of the component masses. In addition, we turned off both stellar 
evolution and the Galactic tidal field, in order to explore the evolution under
the simplest conditions possible. Note that our procedure for fusing binaries
into single stars preserved the total cluster mass, number of stellar systems,
and mass distribution of those systems. In other words, all two-body 
interactions between cluster members were the same as before. The important
difference is that we eliminated the source or sink of energy associated with
the internal motion of binaries.

The results  were both surprising and illuminating. The cluster still undergoes
overall expansion. Figure~13 shows that the central density again falls
steadily. The nominal e-folding time is again 400~Myr, but the actual decline
is not well fit by an exponential. The root cause of the cluster expansion
is that {\it new} binaries continually form and interact with other stars.
This process occurs principally near the relatively dense cluster center, where
the most massive stars reside, along with other, more representative members. 
The component masses in the new binaries are high, typically 
$8\,\langle m\rangle$. Prompt formation of binaries is a well-documented 
occurrence in systems initially containing only single stars \citep{a71}, and
the formation rate is greatly enhanced at higher stellar mass \citep{h75}. In 
our simulations, only 3 or 4 of these systems exist at any time. Nevertheless, 
they are significant dynamically, because of the cluster's relatively low 
gravitational binding.

Any such massive binary with a separation less than 
\hbox{$8\times 10^4\,\,{\rm AU}\,=\,0.4\,\,{\rm pc}$} is hard, i.e., has 
a gravitational potential energy exceeding the initial mean kinetic energy of 
all cluster members. Thus, even the relatively wide binaries formed in these
simulations, with initial separations of order $10^3$~AU, are capable 
of heating the cluster dynamically. As has long been appreciated 
\citep{h75,h83}, the encounter of a hard binary with a third star usually 
results in a harder (tightened) binary. Both this pair and the isolated star 
have more translational kinetic energy than before. The extra energy, which 
comes at the expense of the binary's tightening, is quickly transferred to 
other cluster members.

The same dynamical heating operates, of course, in all stellar groups 
containing binaries. However, very populous systems, such as globular clusters,
have such high gravitational binding that almost all newly formed binaries are 
soft. In this case, energy exchange via three-body encounters has a minor 
effect, and the classical picture of dynamical relaxation via two-body
encounters applies. In relatively sparse systems like open clusters, both
primordial and dynamically formed binaries inject so much energy that they 
impulsively change the velocity distribution function and qualitatively
influence the course of evolution. This stochastic resetting of the velocities,
which was emphasized in the classic study of \citet{t87}, is a conspicuous
feature of the Pleiades evolution, both past and future.

\section{Discussion}

While undertaken primarily to reconstruct the history of the Pleiades, our 
study has shed light on a well-documented, but still poorly understood, feature
of open clusters generally - mass segregation. We demonstrated that the 
currrent, rather high degree of segregation in the Pleiades could not have been
the result of dynamical relaxation from a pristine state with homogeneous mass
distribution. First, the cluster has only been evolving for about half its 
initial relaxation time. Second, a hypothetical cluster starting with no mass 
segregation cannot reach the present level. Quantitatively, the Gini 
coefficient rises, but not enough (recall Figure~9).

Two conclusions may be drawn. The ancient Pleiades must have already had
substantial mass segregation before it drove off the gas. Some other process,
unrelated to dynamical relaxation, must drive this effect, and continues to do
so long into the future (Figure~11). The most obvious candidate is dynamical
friction. A relatively massive star moving through a lower-mass population,
experiences a drag force, causing it sink toward the cluster center. The
associated time scale for braking, $t_{\rm DF}$, can be substantially smaller
than the dynamical relaxation time $t_{\rm relax}$ \citep{s69}. According to 
\citet{pz02}, the quantitative relation is
\begin{equation}
t_{\rm DF} \,=\, 3.3 \ {{\langle m\rangle}\over m}\ t_{\rm relax}\,\,,
\end{equation}
for a heavy star of mass $m$ in a background of average mass 
$\langle m\rangle$.  

\citet{pz02} and other researchers have invoked dynamical friction to explain 
mass segregation, focusing on very populous clusters in which massive star
infall leads to the runaway growth of a central black hole 
\citep[see also][]{g04}. Our work reveals a further, curious aspect of the 
phenomenon. Figures~9 and 11 suggest, and further calculations confirm, that 
$G(t)$ saturates, regardless of its initial value. Why does the degree of mass 
segregation level off? A possible explanation is that, as the most massive 
stars sink to the center, the population there becomes increasingly 
homogeneous. Since $\langle m\rangle/m$ rises, so does $t_{\rm DF}$. In other 
words, mass segregation through dynamical friction may be a self-limiting 
process. 
  
Returning to the prehistory of the Pleiades, another significant finding is
the relatively large size of the initial state. The virial radius $r_v$ began 
at 4~pc, while the projected half-mass radius of the initial cluster was 
about 2~pc. For comparison, the observed half-light radii of embedded clusters,
as seen in the near infrared, range from about 0.5 to 1.0~pc, with some 
outliers on either side \citep{ll03}. Thus, the initial, gas-free 
Pleiades had a radius 2 to 4 times larger than typical embedded systems. It
may plausibly be argued that the Pleiades is an especially populous open 
cluster, and therefore began as a larger configuration, far outside the typical
range. With this caveat in mind, our result suggests that the system expanded 
during its earliest, embedded phase. This swelling, which was accompanied, or 
even preceded, by mass segregation, could have been due to the loss of ambient 
gas during the formation process. Interestingly, observations of
extra-Galactic clusters appear to show a similar, early expansion phase 
\citep[see][and references therein]{ba08}.

We have stressed the importance of binary heating to explain the global
evolution of the Pleiades, both past and future. This is a three-body effect,
not considered in classical studies of dynamical relaxation. As we indicated,
binary heating is more effective in less populous systems, including open
clusters. In the near future, we hope to explore further this general issue of
stellar dynamics, i.e., the demarcation between systems that do and do not 
undergo classical, dynamical relaxation. This study will necessarily delve 
further into the role of binaries. We also intend to repeat our Pleiades 
analysis with another, relatively nearby system of comparable age, to ensure 
that the Pleiades results are representative for the entire class of open 
clusters.  

\acknowledgments
Steve McMillan, one of the authors of Starlab, provided crucial assistance
throughout this project. Not only did he instruct us in the workings of the
code, but he even debugged portions of it at our request. Simon 
Portegies-Zwart, another Starlab author, also gave valued advice. Finally,
we thank James Graham and Chris McKee for their continued interest and 
provocative questions. S.~S. was partially supported by NSF grant AST-0908573. 

\clearpage



\appendix

\section{Physical Scale of the Polytropic Cluster}

As described in the text, the  basic quantities characterizing the initial
state are the polytropic index $n$, along with $N_{\rm tot}$, $r_v$ and $m$.
We show here how to obtain from these the dimensional scale factors
$r_0$, $\rho_0$, and $\Psi_0$. These scale factors, along with the 
dimensionless solution $\psi (\xi)$, allow us to construct the physical cluster
model, as described in the text.

We first define a relative potential energy:
\begin{equation}
W^\prime \,\equiv\,{1\over 2}\,\int_0^{r_t}\!4\,\pi\,r^2\,\rho\,\Psi\,dr \,\,. 
\end{equation}
This has the same form as the true potential energy $W$, but uses the
relative potential $\Psi$ instead of the physical one $\Phi$. Indeed,
solving equation~(3) for $\Phi$ in terms of $\Psi$ and subsituting into
equation~(12) for $W$ yields
\begin{eqnarray}
W \,&=&\, {1\over 2}\,\Phi (r_t)\,M \,-\, W^\prime \\
&=&\,-{{G\,M^2}\over{2\,r_t}}\,-\,W^\prime \,\,.
\end{eqnarray} 
Since $W$ is related to $r_v$, we should next establish a relationship between
$W^\prime$ and other dimensional quantities.

Following \citet{k66}, we define a nondimensional potential as
\begin{eqnarray}
\beta \,&\equiv&\, \int_0^{\xi_t}\!4\,\pi\,\xi^2\,\psi^{n+1}\,d\xi \\
&=&\, {1\over r_0^3}\,\int_0^{r_t}\!4\,\pi\,r^2\,{\rho\over \rho_0}\,
{\Psi\over\Psi_0}\,dr \nonumber \\
&=&\,{{2\,W^\prime}\over{\rho_0\,r_0^3\,\Psi_0}} \,\,.
\end{eqnarray}
We also define a nondimensional cluster mass:
\begin{eqnarray}
\mu \,&\equiv&\, \int_0^{\xi_t}\!4\,\pi\,\xi^2\,\psi^n\,d\xi \\
&=&\,{1\over r_0^3}\,\int_0^{r_t}\!4\,\pi\,r^2\,{\rho\over \rho_0} \,dr 
\nonumber \\
&=&\,{M\over{\rho_0\,r_0^3}} \,\,.
\end{eqnarray}
Both $\beta$ and $\mu$ can be calculated using the solution $\psi (\xi)$. Their
ratio is
\begin{eqnarray}
{\beta\over\mu} \,&=&\,{{2\,W^\prime}\over{M\,\Psi_0}} \\
&=&\,{{2\,W^\prime\,r_0\,\mu}\over{4\,\pi\,G\,M^2}} \,\,,
\end{eqnarray}
where we have used equations~(A7) and (9) to make the last transformation. 
Solving this equation for $W^\prime$ and substituting into equation~(A3), we 
find
\begin{equation}
W \,=\, -{{G\,M^2}\over 2}\,
\left({1\over r_t} \,+\, {{4\,\pi\,\beta}\over{r_0\,\mu^2}}\right) \,\,.
\end{equation}

This last relation gives us more information about the virial radius. We now 
see that the nondimensional version, \hbox{$\xi_v\,\equiv\,r_v/r_0$}, obeys
\begin{equation}
{1\over\xi_v} \,=\, {1\over\xi_t} \,+\,{{4\,\pi\,\beta}\over\mu^2} \,\,.
\end{equation}
Thus, $\xi_v$ can be obtained at once from the nondimensional solution. Since
$r_v$ itself is an input, the dimensional scale radius, $r_0$, can be
obtained from
\begin{equation}
r_0 \,=\, {r_v\over\xi_v} \,\,.
\end{equation}
Similarly, the central density $\rho_0$ is
\begin{eqnarray}
\rho_0 \,&=&\, {M\over{\mu\,r_0^3}} \\
&=&\, \left({\xi_v^3\over\mu}\right){{N_{\rm tot}\,m}\over r_v^3} \,\,.      
\end{eqnarray}
while $\Psi_0$ is found from
\begin{eqnarray}
\Psi_0 \,&=&\, 4\,\pi\,G\,\rho_0\,r_0^2 \\
&=&\, \left({{4\,\pi\,\xi_v}\over\mu }\right) 
{{G\,N_{\rm tot}\,m}\over r_v} \,\,.
\end{eqnarray}

\clearpage


\section{Distribution of Component Masses within Binaries}

Let $\phi$ be the normalized distribution of single star masses. In the text, 
we used the same functional notation when referring to the mass distribution
for the {\it initial} cluster (see eq.~(17)). We are now concerned with the
{\it evolved} cluster. Our assumed functional form will be different, but we
retain the notation for simplicity. We again let $\gamma$ be the binary mass
correlation parameter, as in equation~(20) of the text. Here we show how to
find, for a given $\gamma$, the distribution of primary and secondary masses, 
as well as the distribution of the secondary-to-primary mass ratio $q$. Our
final expressions for the various distributions are rather cumbersome and
not especially illuminating; we therefore limit ourselves to outlining the 
derivation for a generic single-star function $\phi$. 

We first let the primary and secondary masses have provisional masses 
$m_p^\ast$ and $m_s^\ast$, respectively. Assume that both components within
binaries are drawn independently from the same distribution $\phi$. Then the
two-dimensional mass function of the binaries is
\begin{equation}
\Phi_b \left(m_p^\ast, m_s^\ast\right) \,=\, 2\,\phi (m_p^\ast)\,
\phi (m_s^\ast) \,\,.
\end{equation} 
Here,
$\Phi_b \left(m_p^\ast, m_s^\ast\right)\,\Delta m_p^\ast\,\Delta m_s^\ast$ 
is the probability of finding a system with primary mass between $m_p^\ast$
and $m_p^\ast\,+\,\Delta m_p^\ast$, and secondary mass between $m_s^\ast$ and
$m_s^\ast\,+\,\Delta m_s^\ast$. This function is normalized so that
\begin{equation}
\int_{m_{\rm min}}^{m_{\rm max}}\!dm_p^\ast
\int_{m_{\rm min}}^{m_p^\ast}\!dm_s^\ast\,\,\Phi_b ({m_p^\ast} ,{m_s^\ast})
\,=\,1 \,\,.
\end{equation}  
As explained in \S 2.2 of Paper~I, the initial factor of 2 on the righthand 
side of equation~(B1) accounts for the different integration limits of 
$\Phi_b$ and $\phi$ (for the latter, see eq.~(19)).

To implement binary mass correlation, we consider new primary and secondary
masses, $m_p$ and $m_s$, related to the previous ones by
\begin{eqnarray}
m_p &=& m_p^\ast \\
m_s &=& m_s^\ast \left({{m_p^\ast}\over{m_s^\ast}}\right)^\gamma \,\,.
\end{eqnarray} 
We are interested in the distribution function \hbox{$\Phi_b (m_p, m_s)$}, 
which is 
\begin{equation}
\Phi_b (m_p, m_s) \,=\, \Phi_b (m_p^\ast, m_s^\ast)\,\,
\left|{{\partial\,(m_p^\ast , m_s^\ast)}\over
{\partial\,(m_p , m_s )}}\right| \,\,.
\end{equation}
After evaluating the Jacobian, we find
\begin{equation}
\Phi_b (m_p, m_s)\,=\, {2\over{1-\gamma}}
\left({m_s\over m_p}\right)^{\gamma/(1-\gamma)}\!\phi (m_p)\,\,\,
\phi\!\left[m_s\left({m_s\over m_p}\right)^{\gamma/(1-\gamma)}\right] \,\,.
\end{equation} 

Let us first consider $\phi_p (m_p)$, the distribution of primary masses. This
function is the integral of $\Phi_b (m_p, m_s)$ over all appropriate values of
$m_s$:
\begin{equation}
\phi_p (m_p) \,=\, \int_{m_{\rm s,min}}^{m_{\rm s,max}}\!
\Phi_b (m_p, m_s)\,dm_s \,\,.
\end{equation}
The largest mass a secondary can have, given the primary mass, is $m_p$ itself:
\begin{equation}
m_{\rm s,max} \,=\, m_p \,\,.
\end{equation}
However, the smallest mass is {\it not} $m_{\rm min}$. This is indeed the 
smallest mass for $m_s^\ast$. The correlation of primary and secondary masses 
implies that the minimum for $m_s$ is 
\begin{equation}
m_{\rm s,min} \,=\, m_{\rm min} 
\left({m_p\over{m_{\rm min}}}\right)^\gamma \,\,.
\end{equation}
Thus, the two integration limits in equation~(B7) are themselves functions of
$m_p$.

The secondary mass function, $\phi_s (m_s)$, is similarly found by integrating
\hbox{$\Phi_b (m_p, m_s)$} over all possible primary masses:
\begin{equation}
\phi_s (m_s) \,=\, \int_{m_{\rm p,min}}^{m_{\rm p,max}}\!
\Phi_b (m_p, m_s)\,dm_p \,\,.
\end{equation}
The smallest value a primary mass can be, for a given secondary, is $m_s$:
\begin{equation}
m_{\rm p,min} \,=\, m_s \,\,. 
\end{equation}
Somewhat surprisingly, the largest value is not necessarily $m_{\rm max}$, 
again because of the imposed correlation. To find the correct maximum, we solve
equation~(B4) for $m_p^\ast$:
\begin{equation}
m_p^\ast \,=\, m_p \,=\, m_s^{1/\gamma}\,
\left(m_s^\ast\right)^{(\gamma - 1)/\gamma} \,\,.
\end{equation}
Since $\gamma$ lies between 0 and 1, the exponent of $m_s^\ast$ is negative.
Thus, for a given $m_s$, $m_p$ is greatest when $m_s^\ast$ is smallest. Since
the lowest value of $m_s^\ast$ is $m_{\rm min}$, we have
\begin{eqnarray}
m_{\rm p,max} \,&=&\, m_s^{1/\gamma}
\left(m_{\rm min}\right)^{{(\gamma - 1)}/\gamma} \\
&=&\, m_{\rm min}
\left({m_s\over{m_{\rm min}}}\right)^{1/\gamma} \,\,. 
\end{eqnarray}
However, for \hbox{$m_s > m_{\rm min} (m_{\rm max}/m_{\rm min})^\gamma$}, this
equation says that \hbox{$m_{\rm p,max} > m_{\rm max}$}, which is impossible.
In summary, $m_{\rm p,max}$ is given by 
\begin{equation}
m_{\rm p,max} \,=\, \left\{
\begin{array}{rl}
m_{\rm min} \left(m_s/m_{\rm max}\right)^{1/\gamma} 
& \,\, m_s  \,\le\,m_{\rm min} 
\left(m_{\rm max}/m_{\rm min}\right)^\gamma 
\\
 m_{\rm max} & \,\, m_s \,>\,m_{\rm min}
\left(m_{\rm max}/m_{\rm min}\right)^\gamma 
\,\,.
\end{array} \right.
\end{equation}

Finally, we need $\phi_q (q)$, the distribution of the binary mass ratio
\hbox{$q\,\equiv\, m_s/m_p$}. As a first step, we find the two-dimensional 
mass function \hbox{$\Phi_b (m_p, q)$}. Proceeding as before, we have
\begin{eqnarray}
\Phi_b (m_p, q) \,&=&\, \Phi_b (m_p^\ast, m_s^\ast)\,\,
\left|{{\partial\,(m_p, m_s)}\over
{\partial\,(m_p , q)}}\right| \\
&=&\, m_p\,\,\Phi_b (m_p , m_s) \\
&=&\, {{2\,m_p}\over{1-\gamma}}\,\,q^{\gamma/(1-\gamma)}\,\phi (m_p)\,\,
\phi\!\left[m_p\,q^{1/(1-\gamma)}\right] \,\,.
\end{eqnarray}
The desired distribution is the integral of \hbox{$\Phi_b (m_p, q)$} over 
suitable $m_p$-values:
\begin{equation}
\phi_q (q) \,=\, \int_{m_{\rm p,min(q)}}^{m_{\rm p,max(q)}}\!
\Phi_b (m_p, q)\,dm_p \,\,.
\end{equation}
As the notation indicates, the limits of $m_p$ are subject to the restriction 
of a fixed $q$. Now $q$ itself is given in terms of 
\hbox{$m_p^\ast\,(=\,m_p)$} and $m_s^\ast$ by
\begin{eqnarray}
q \,&=&\, {m_s^\ast\over m_p}
\left({m_p\over m_s^\ast}\right)^\gamma \\
&=&\, \left({m_p\over m_s^\ast}\right)^{\gamma-1} \,\,.
\end{eqnarray}
Solving the last equation for $m_p$ gives
\begin{equation}
m_p \,=\, m_s^\ast\,\, q^{-1/(1-\gamma)} \,\,.
\end{equation}
Since the exponent of $q$ is negative, and since $q$ itself lies between 0 and
1, we see that \hbox{$m_p > m_s^\ast$}. Thus, for any $q$-value, there is 
always some $m_s^\ast$ for which \hbox{$m_p \,=\, m_{\rm max}$}. We therefore 
set
\begin{equation}
m_{\rm p,max(q)} \,=\, m_{\rm max} \,\,.
\end{equation}
The smallest value of $m_p$ corresponds to \hbox{$m_s^\ast\,=\,m_{\rm min}$}. 
It follows that
\begin{equation}
m_{\rm p,min(q)} \,=\, m_{\rm min}\,\, q^{-1/(1-\gamma)} \,\,.
\end{equation}

We may, in principle, perform the integrals in equations~(B7), (B10), and (B19)
for any specified single-star function $\phi (m)$. In practice, we choose a 
lognormal:
\begin{equation}
\phi (m) \,=\, {D\over m}\,{\rm exp}\,\left(-y^2\right) \,\,,
\end{equation}
where $D$ is the normalization constant, and the variable $y$ is given by
equation~(18) in the text. With this form of $\phi (m)$, the integrations may
all be done analytically, although we do not reproduce the rather lengthy 
results here.

\begin{figure}
\plotone{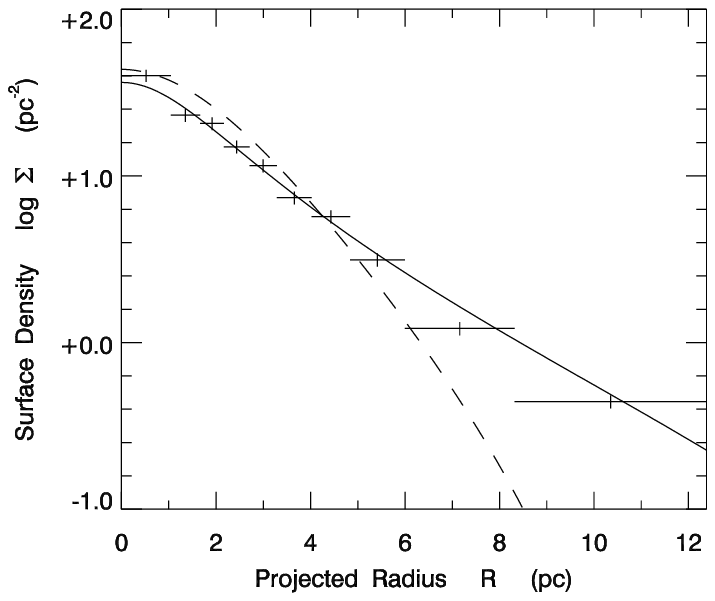}
\caption{Surface number density as a function of projected radius. The dashed
curve represents our initial configuration, an {$n\,=\,3$} polytrope. The
solid curve is a \citet{k62} model fit to our simulation results for the 
evolved cluster. Table 2 lists the parameters for this optimal model. The
numerical results displayed are an average of 25 simulation runs. Also shown 
are Pleiades data with error bars, taken from Paper~I.}
\end{figure}

\clearpage

\begin{figure}
\plotone{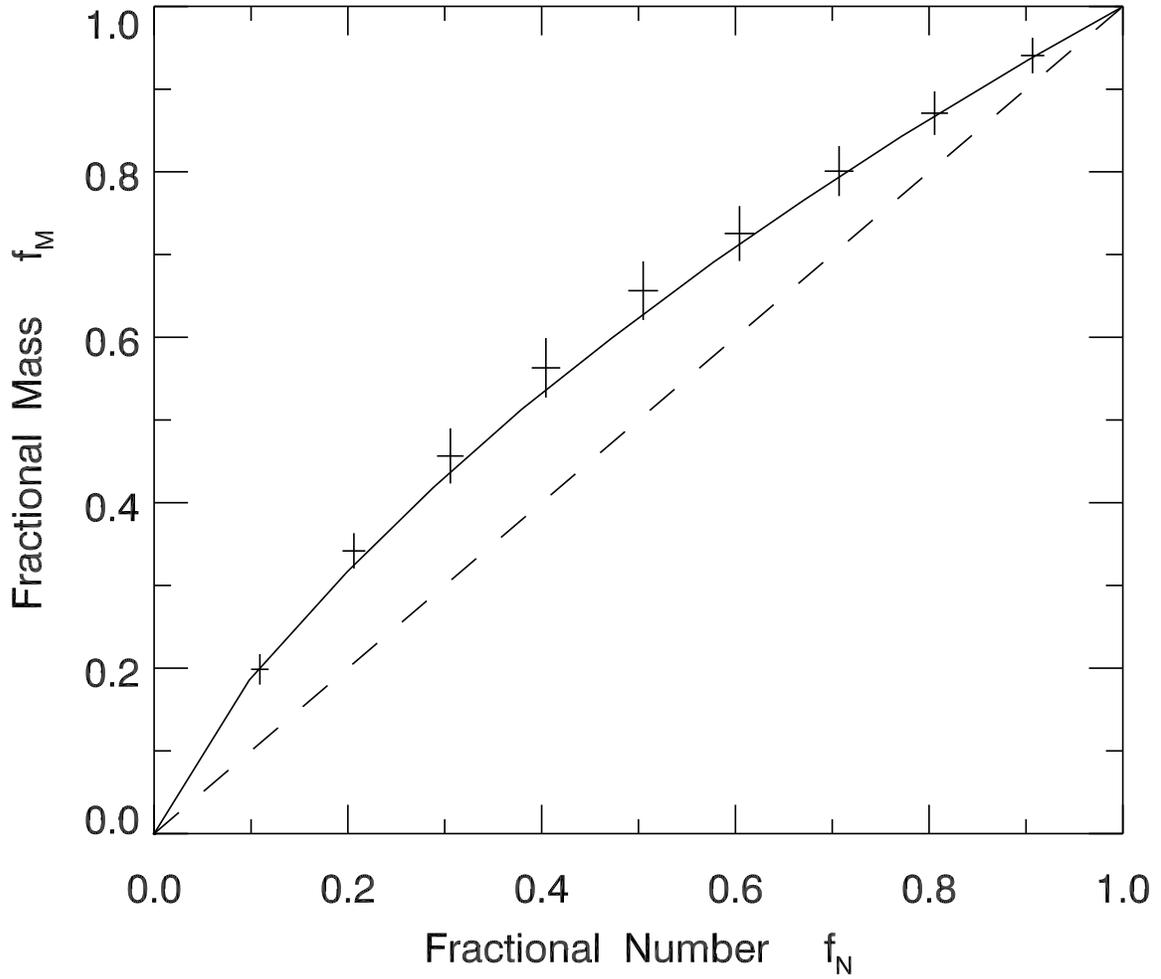}
\caption{Fractional mass versus fractional number for the Pleiades. The solid
curve shows the average results of our simulations. The crosses represent
Pleiades data with error bars, taken from Paper~I. The dashed diagonal line is
the hypothetical result for zero mass segregation.} 
\end{figure}

\clearpage

\begin{figure}
\plotone{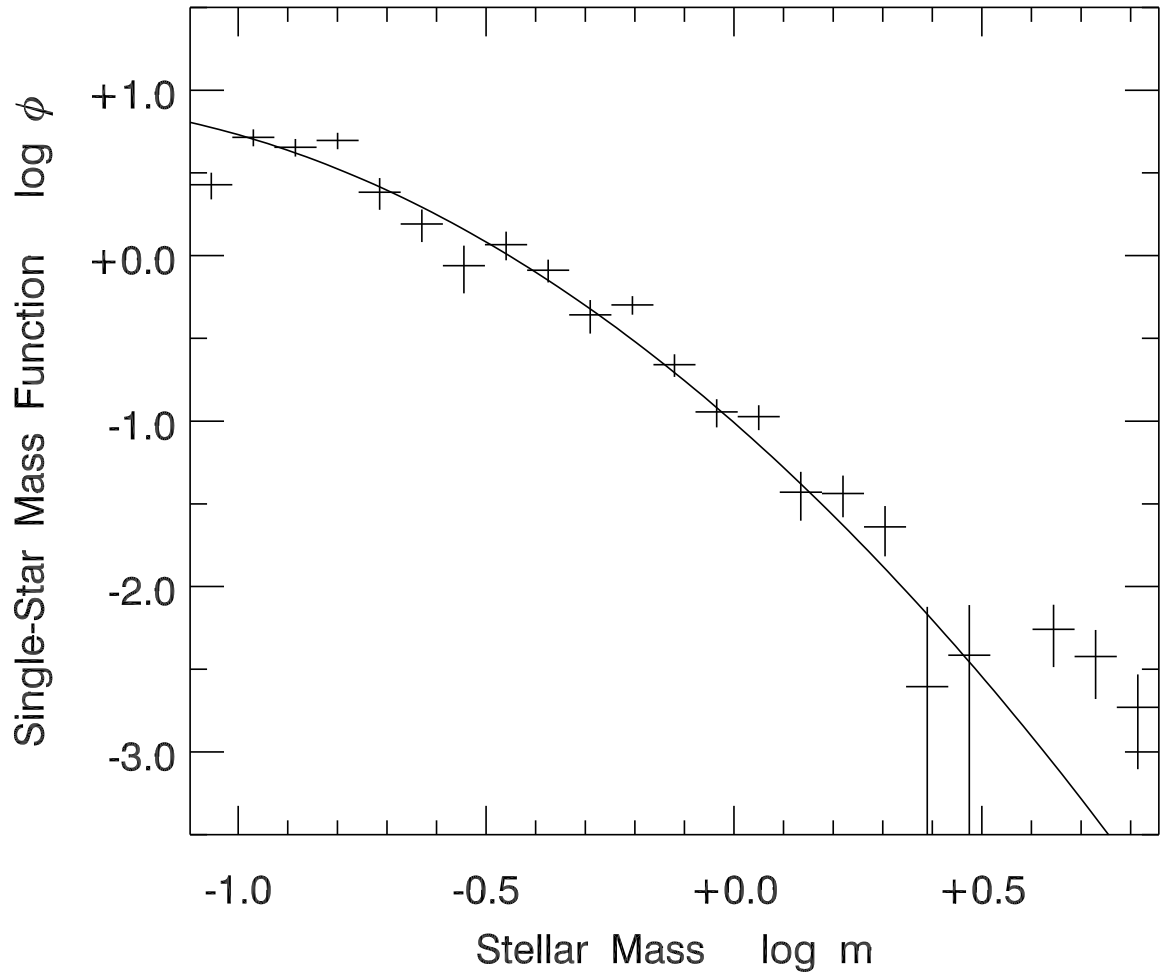}
\caption{Single-star mass function for the evolved cluster. The solid curve is
a lognormal fit to simulation data. Also shown are Pleiades data, with error
bars, from Paper~I.} 
\end{figure}

\clearpage

\begin{figure}
\plotone{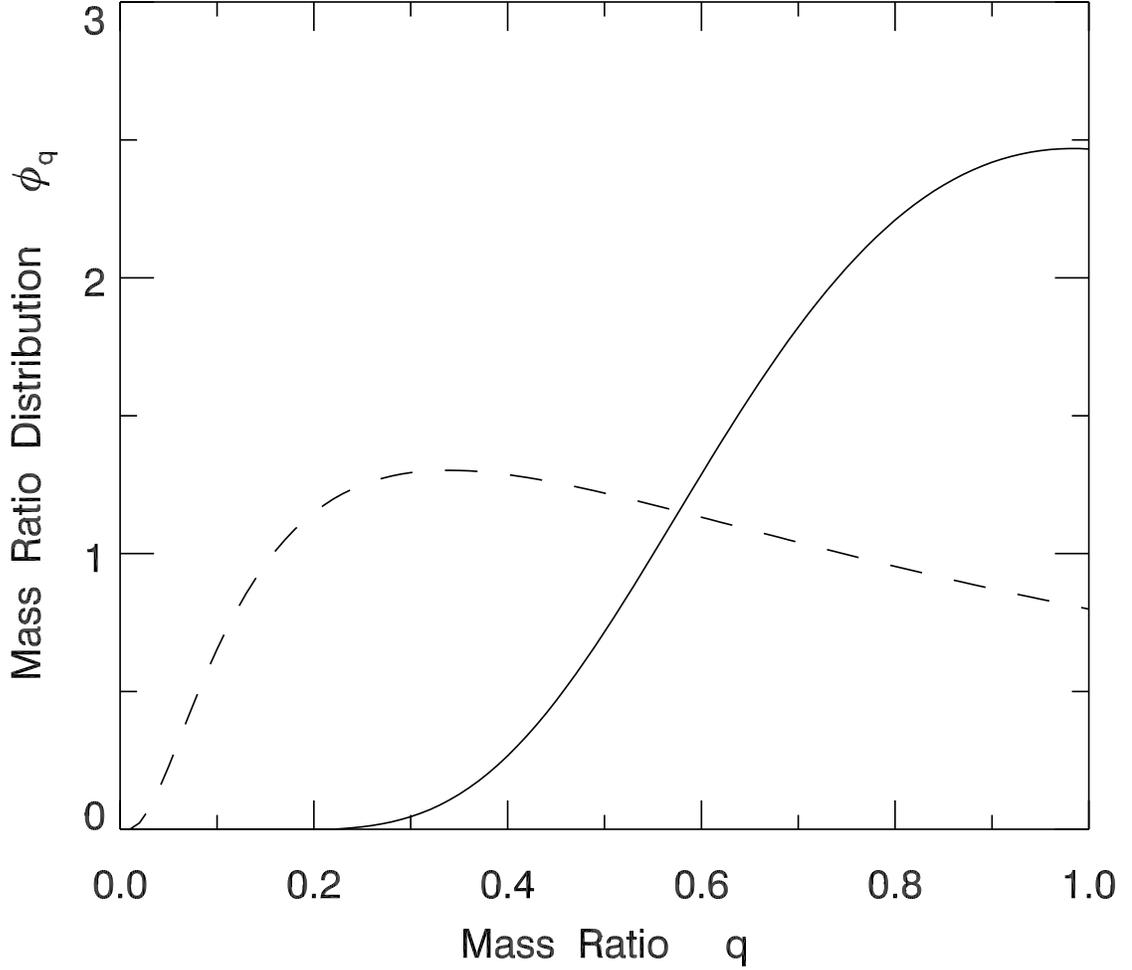}
\caption{Initial distribution of the mass ratio within binaries,
\hbox{$q\,\equiv\,m_s/m_p$}. The solid curve was obtained using a lognormal 
fit to the calculated single-star mass function, including the proper 
binary mass correlation parameter $\gamma$. The dashed curve is the 
hypothetical distribution obtained with the same single-star mass function, but
with no mass correlation (\hbox{$\gamma\,=\,0$}).} 
\end{figure}

\clearpage

\begin{figure}
\plotone{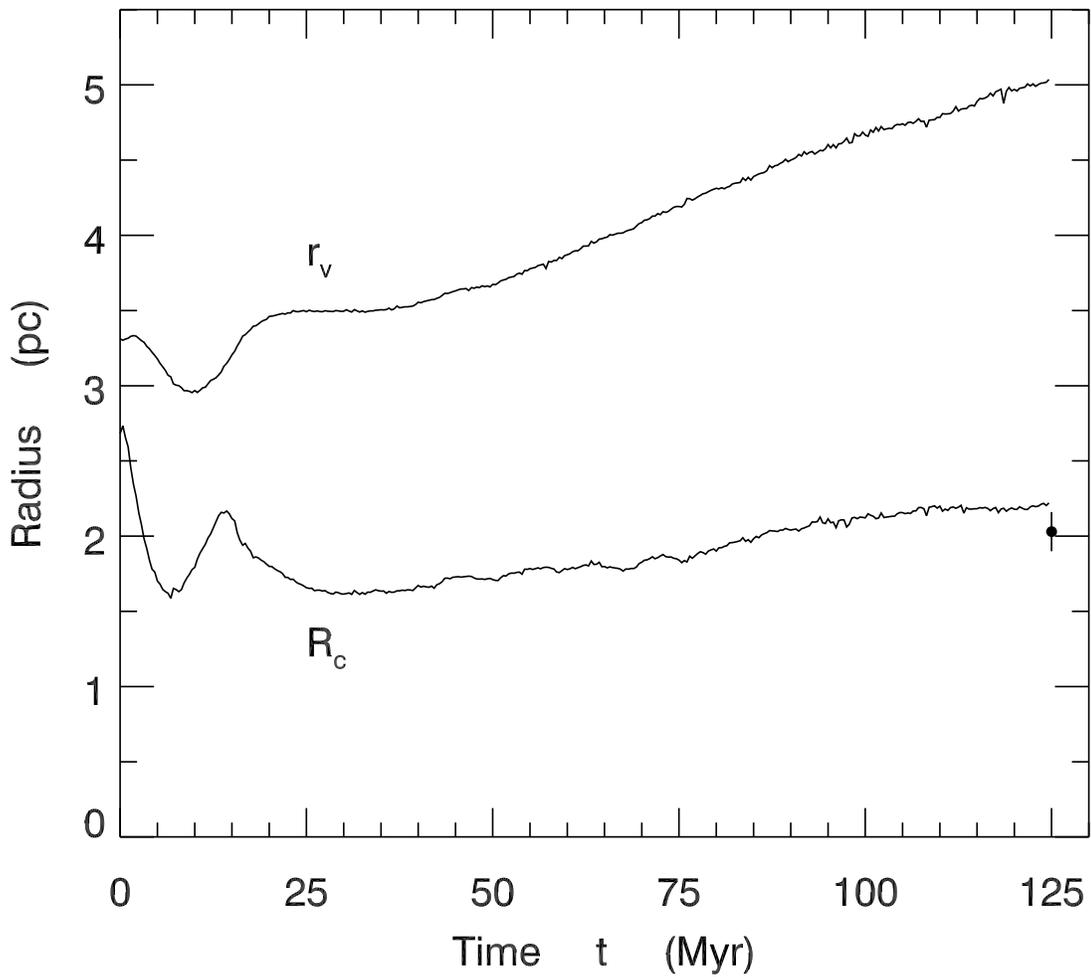}
\caption{Evolution of characteristic radii. The upper curve shows the 
three-dimensional virial radius $r_v$, and is an average over simulation 
runs. The lower curve shows the projected core radius $R_c$, and is also
an average. The data point in the lower right is the observed Pleiades value 
for $R_c$, along with error bars.} 
\end{figure}

\clearpage

\begin{figure}
\plotone{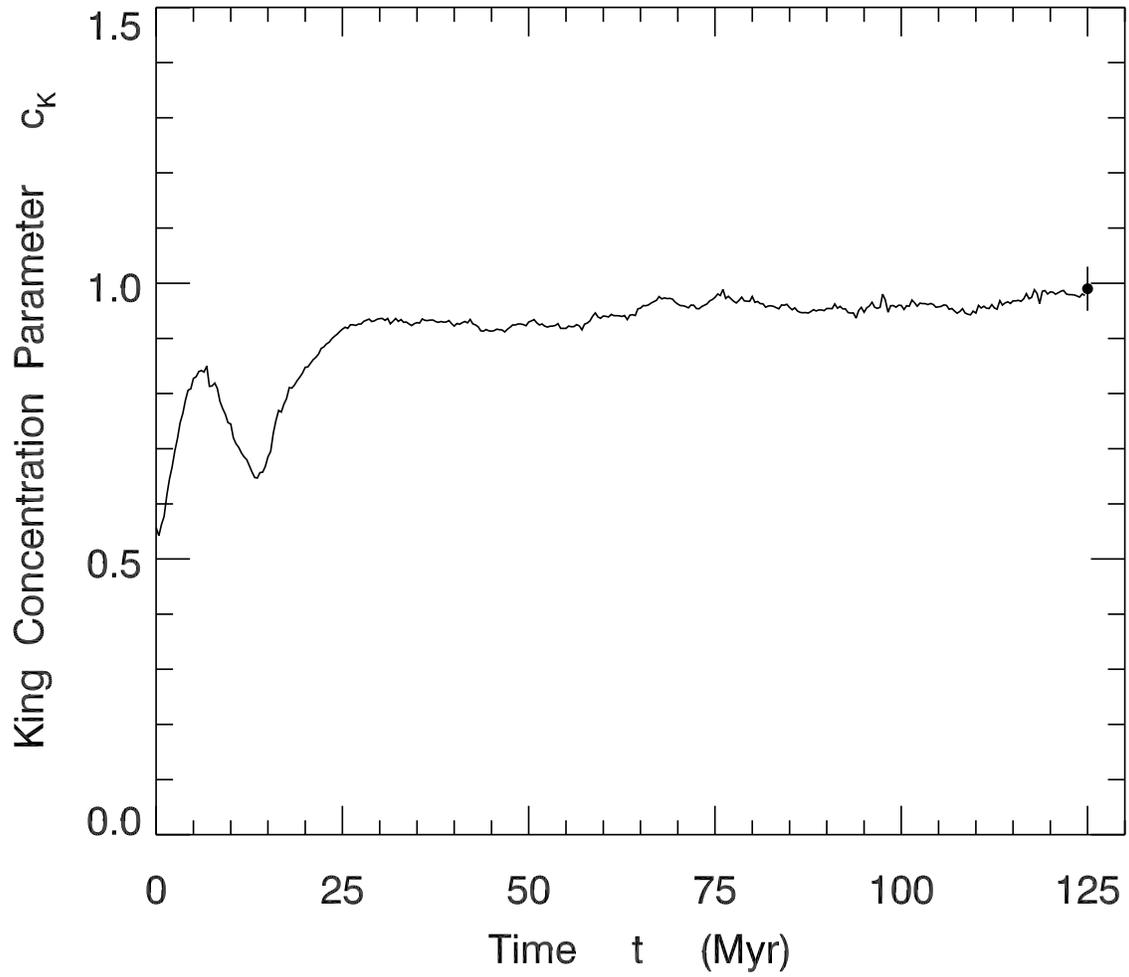}
\caption{Evolution of the King concentration parameter. The curve is an
average over simulation runs. The data point is the observed Pleiades value.} 
\end{figure}

\clearpage

\begin{figure}
\plotone{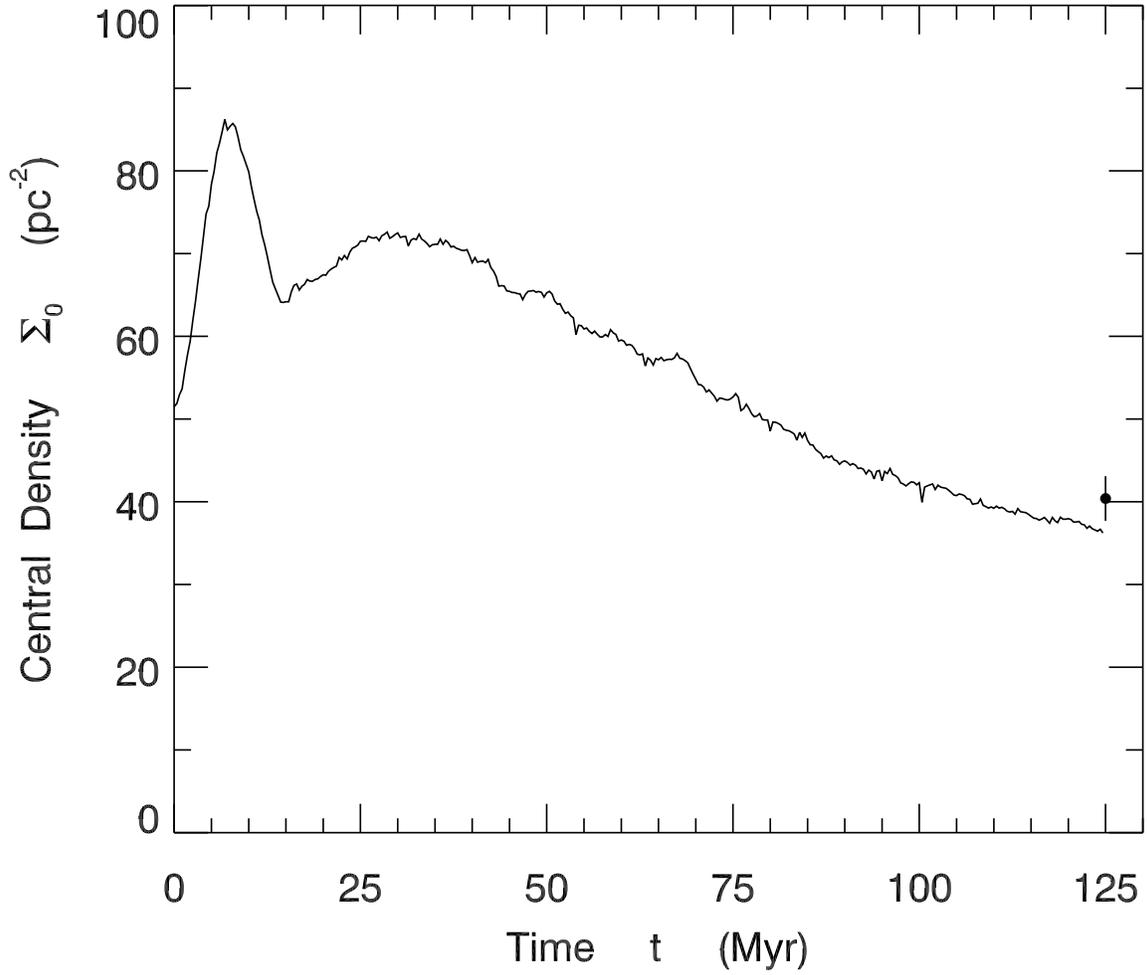}
\caption{Evolution of the central surface number density. Shown is the average
over simulation results. The observed Pleiades value is represented by the 
data point.} 
\end{figure}

\clearpage

\begin{figure}
\plotone{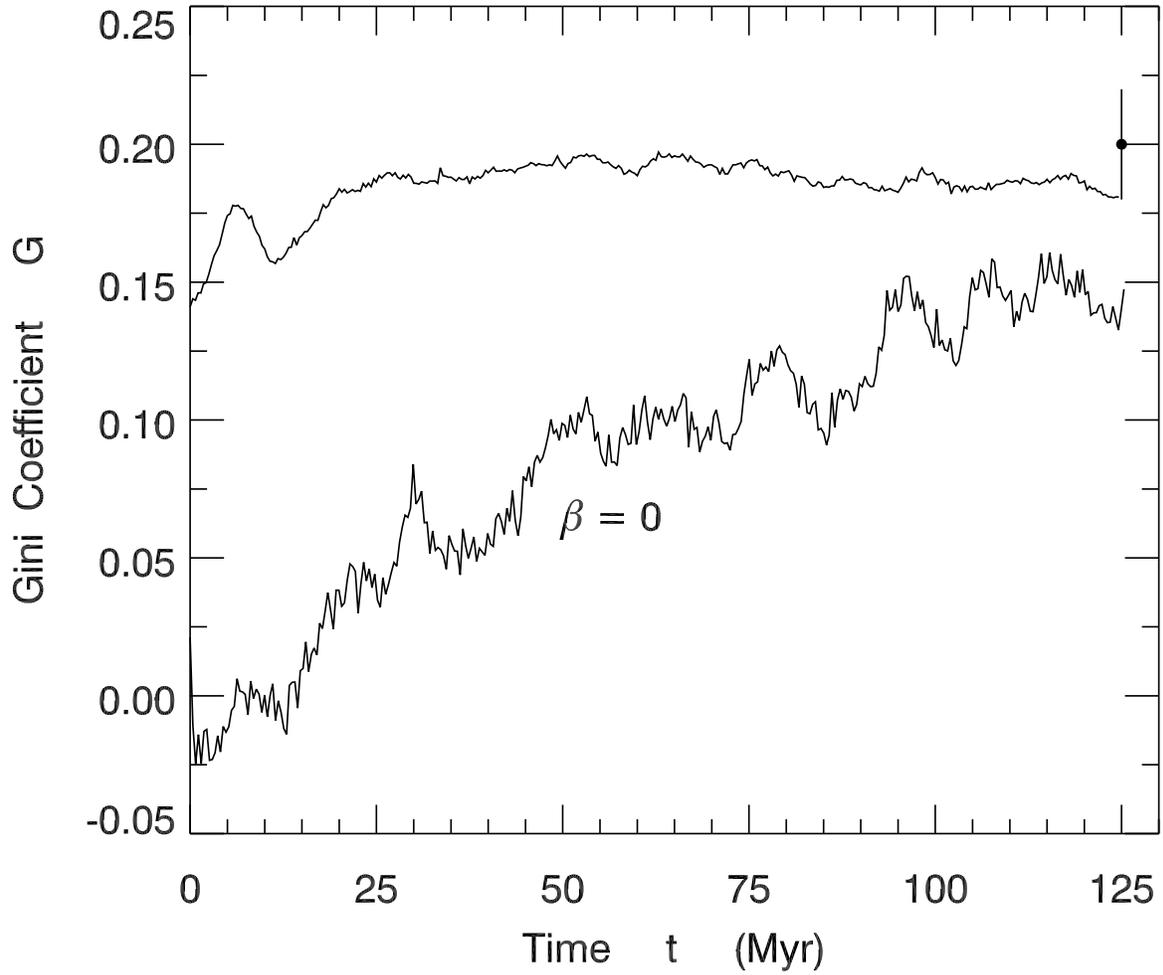}
\caption{Evolution of the Gini coefficient. The upper curve is an average over
simulation results. To the right of this curve is the observed Pleiades value.
The lower curve shows the result from a single simulation run in which the
mass segregation parameter $\beta$ was artificially set to zero.} 
\end{figure}

\begin{figure}
\plotone{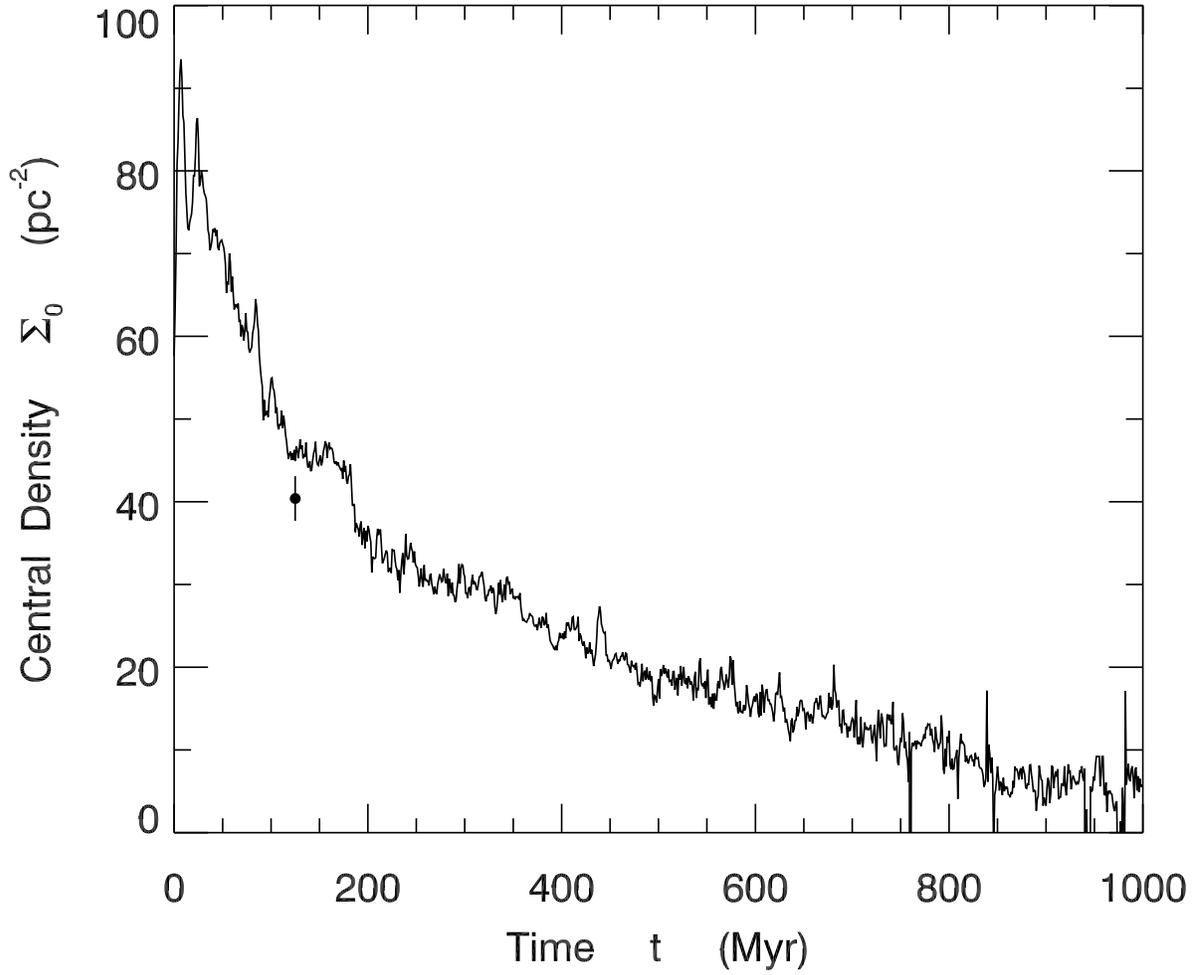}
\caption{Evolution of the central surface number density over a total time
of 1~Gyr. The data point is the present-day Pleiades value, with errors
indicated.}
\end{figure}

\begin{figure}
\plotone{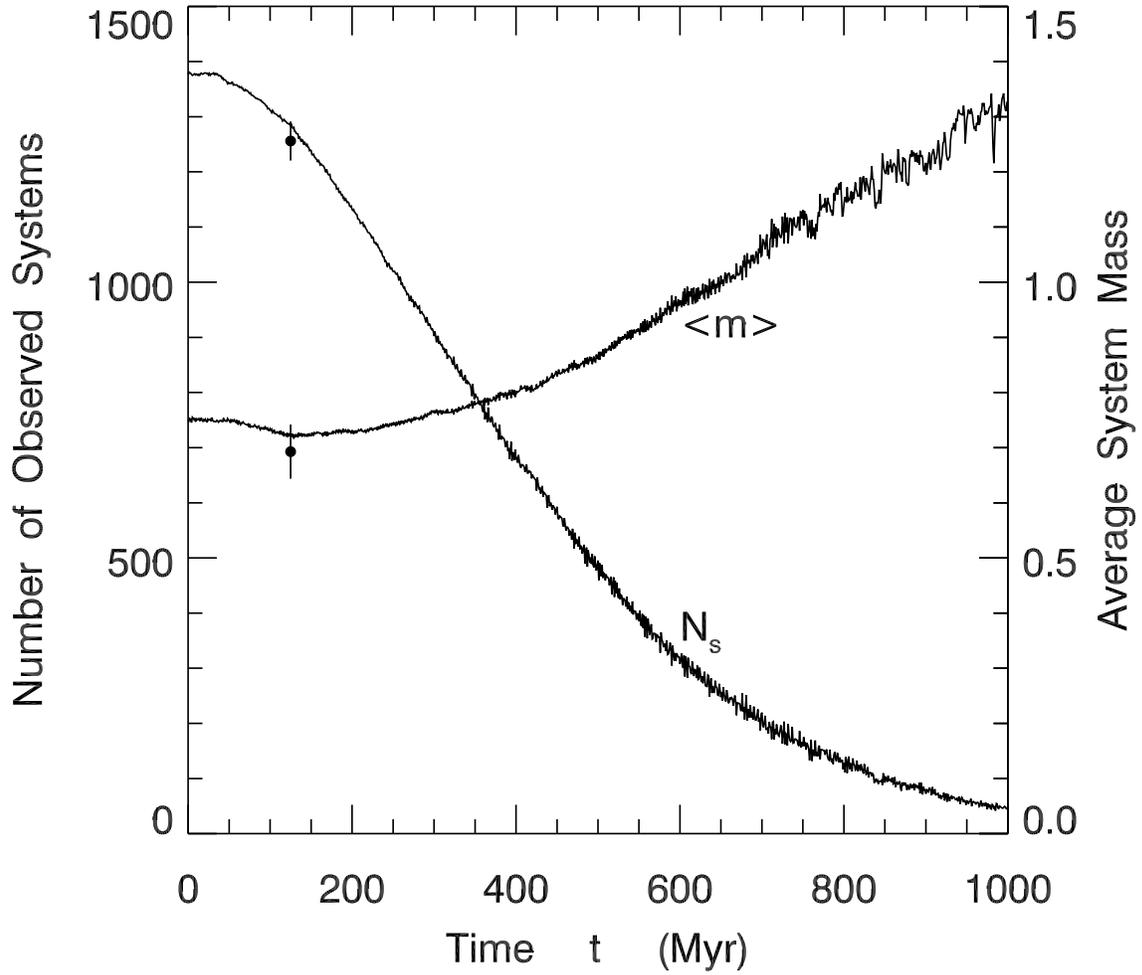}
\caption{Evolution of the total number of stellar systems, $N_s$, and the
average system mass $\langle m\rangle$, over 1~Gyr. Both quantities refer to
systems within the initial Jacobi radius of 14.4~pc. The data points show
the current Pleiades values, with error bars.}
\end{figure}

\begin{figure}
\plotone{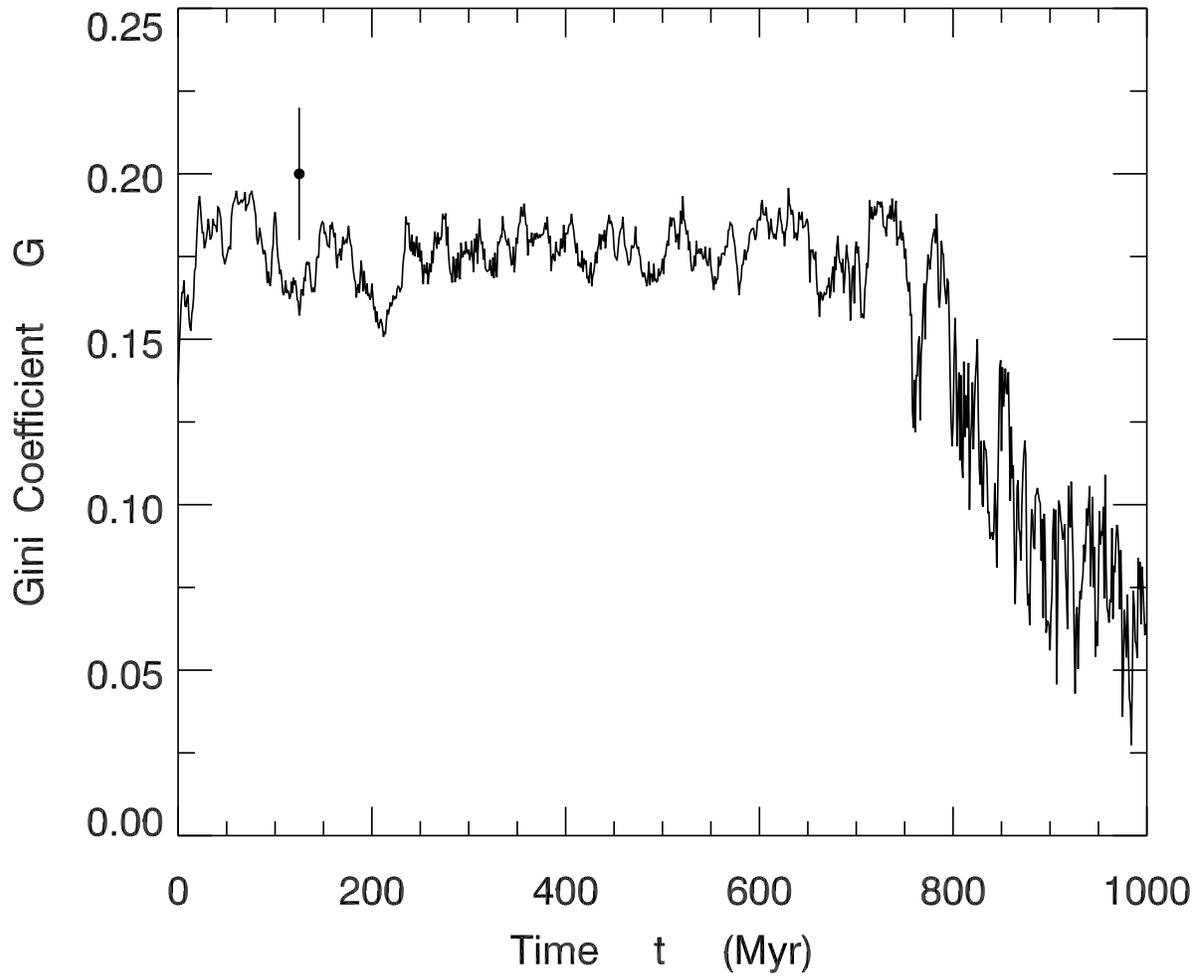}
\caption{Evolution of the Gini coefficient over 1~Gyr. Note the large scatter
at late times, reflecting the falloff in total cluster population. The data
point to the left is the current Pleiades value, along with error bars.}
\end{figure}

\begin{figure}
\plotone{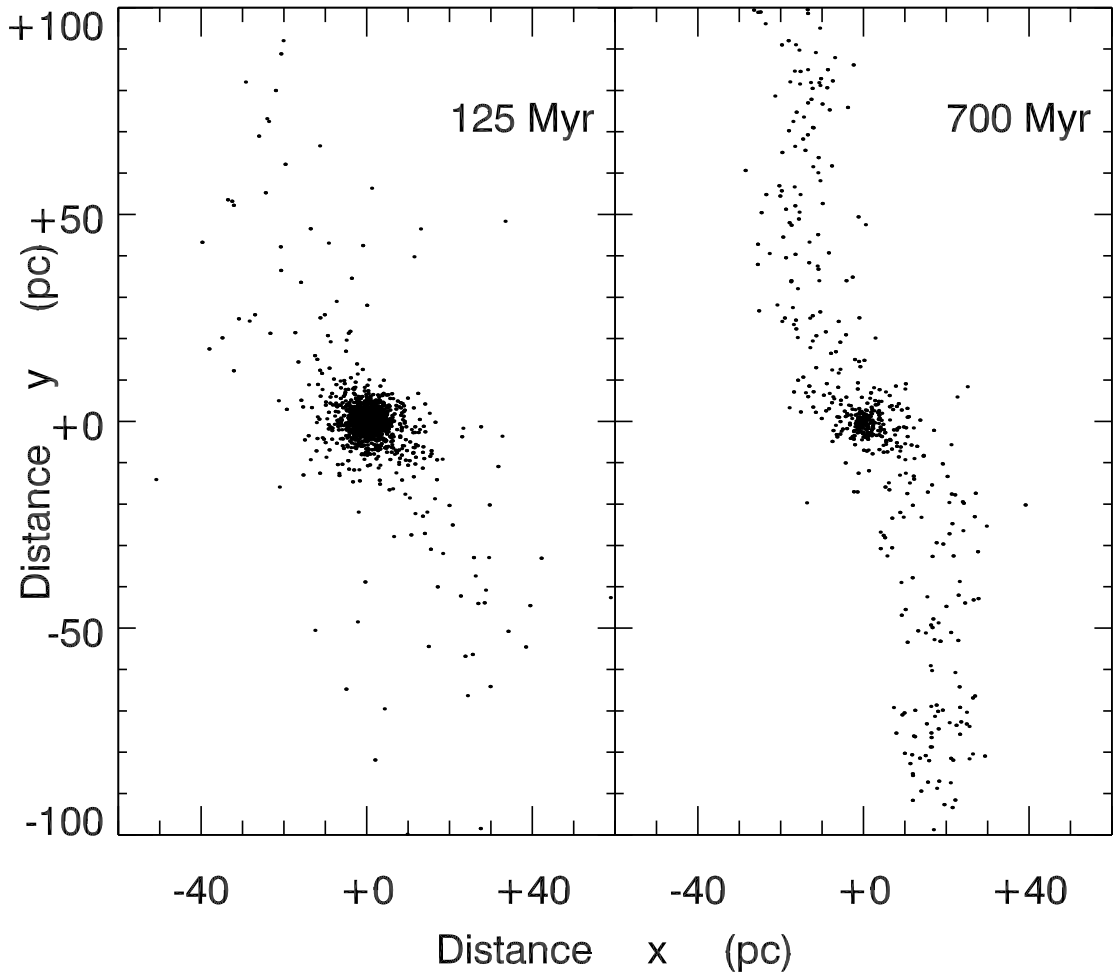}
\caption{Positions of Pleiades members projected onto the Galactic plane. The 
data are from a single, representative simulation, for the two epochs 
indicated. A terrestrial observer is located 133~pc in the negative 
$x$-direction. The Galactic Center is in the same direction, but 8~kpc 
distant. Galactic rotation is in the positive $y$-direction.} 
\end{figure}

\begin{figure}
\plotone{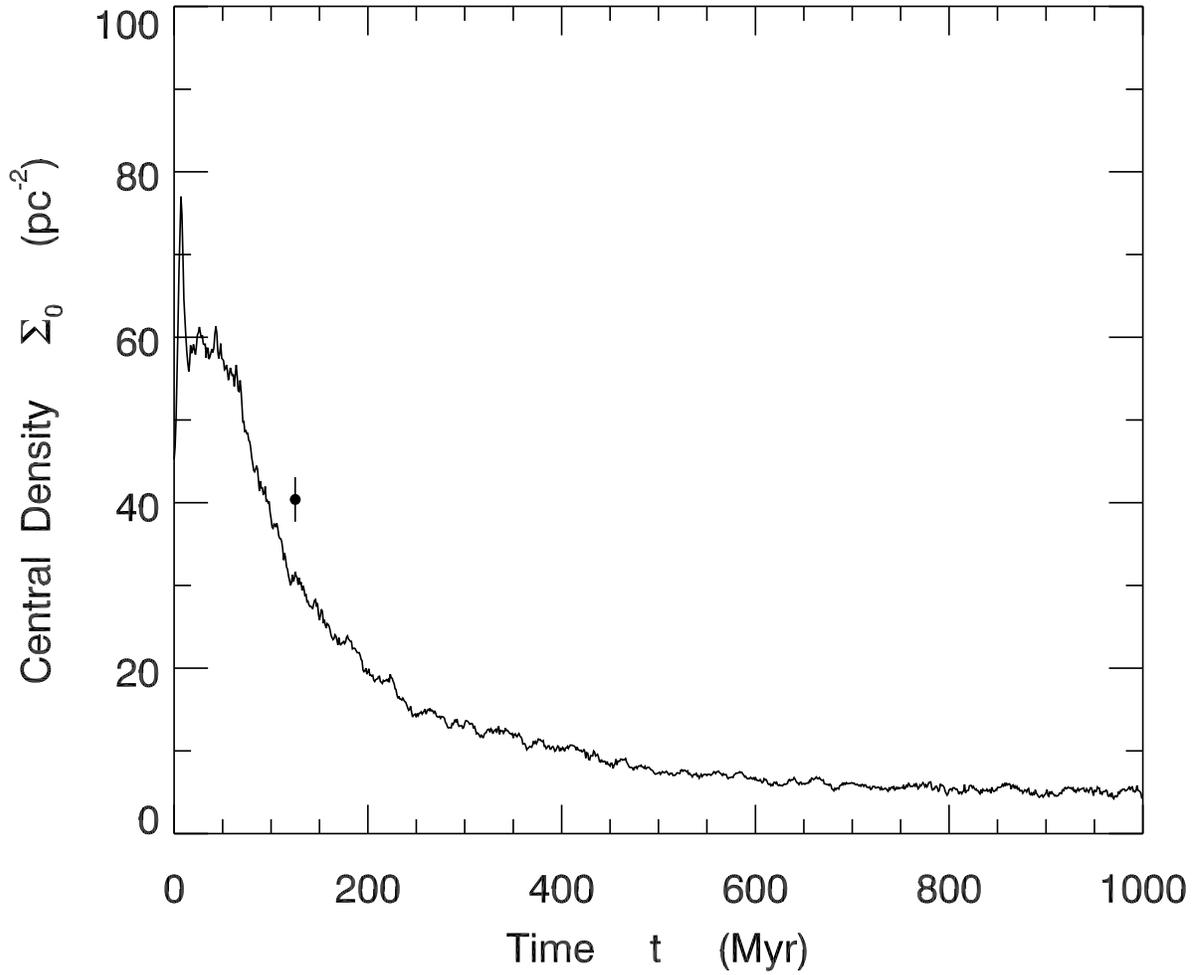}
\caption{Long-term evolution of the central number surface density for a 
cluster with no primordial binaries. The curve was obtained by averaging 9
simulation runs. The data point shows the current Pleiades central density, 
with errors.} 
\end{figure}

\clearpage

\begin{center}
\begin{table}[t]
\caption{Initial Cluster Parameters}
\begin{tabular}{cll}
\tableline\tableline
Symbol & Definition & Optimal Value\\ 
\tableline
$n$ &polytropic index &$3.0 \pm 1.3$\\
$N_{\rm tot}$ &number of stellar systems &$1215 \pm 59$\\
$r_v$ &virial radius &$4.0 \pm 0.9$ pc\\
$m_0$ &centroid of mass function &$0.12 \pm 0.04\,\,\Msun$ \\
$\sigma_m$ & width of mass function &$0.33 \pm 0.06$ \\
$\alpha$ &exponent in mass function &$-2.20 \pm 0.04$ \\
$b$ &fraction of binaries &$0.95 \pm 0.08$ \\
$\gamma$ &mass correlation in binaries &$0.73 \pm 0.09$\\
$\beta$ &degree of mass segregation &$0.5 \pm 0.3$ \\
\tableline
\end{tabular}
\end{table}
\end{center}

\clearpage

\begin{center}
\begin{table}[t]
\caption{Evolved Cluster Properties}
\begin{tabular}{clll}
\tableline\tableline
Symbol & Definition & Calculated Value & Pleiades Value\\ 
\tableline
$N_{\rm s}$ &number of point sources &$1244 \pm 32$ &$1256 \pm 35$\\
$N_4$ &number of systems with $m > 4$ &$13 \pm 4$ &$11 \pm 3$\\
$M_{\rm tot}$ &cluster mass &$939 \pm 30\,\,\Msun$ &$870 \pm 35\,\,\Msun$\\
$b_{\rm unres}$ &unresolved binary fraction &$0.68 \pm 0.02$ 
&$0.68 \pm 0.02$\\
$m_0$ &centroid of mass function &$0.12 \pm 0.03\,\,\Msun$ 
&$0.14 \pm 0.05\,\,\Msun$ \\
$\sigma_m$ & width of mass function &$0.49 \pm 0.05$ 
& $0.46 \pm 0.04$\\
$\gamma$ &binary correlation index & $0.66 \pm 0.01$ &$0.65 \pm 0.05$\\
$R_c$ &core radius &$2.2 \pm 0.4\,\,{\rm pc}$ &$2.0 \pm 0.1\,\,{\rm pc}$\\
$c_K$ &King concentration parameter &$0.98 \pm 0.09$ &$0.99 \pm 0.04$\\
$\Sigma_0$ &central surface density &$36 \pm 8\,\,{\rm pc}^{-2}$ 
&$40 \pm 3\,\,{\rm pc}^{-2}$\\
$G$ &Gini coefficient &$0.18 \pm 0.02$ &$0.20 \pm 0.02$\\
\tableline
\end{tabular}
\end{table}
\end{center}

\end{document}